\documentclass[fleqn,usenatbib]{mnras}

\usepackage{newtxtext,newtxmath}

\usepackage[T1]{fontenc}
\usepackage[table]{xcolor}
\usepackage[dvipsnames]{}

\DeclareRobustCommand{\VAN}[3]{#2}
\let\VANthebibliography\thebibliography
\def\thebibliography{\DeclareRobustCommand{\VAN}[3]{##3}\VANthebibliography}

\usepackage{graphicx}	
\usepackage{amsmath}	

\title[Planet multiplicity and WD pollution]{Disentangling the parameter space: The role of planet multiplicity in triggering dynamical instabilities on planetary systems around white dwarfs}

\author[R. F. Maldonado et al.]{
R. F. Maldonado,$^{1}$, E. Villaver $^{2}$, A. J. Mustill $^{3}$, M. Ch\'avez $^{1}$  \\
\thanks{E-mail: raulfms@inaoep.mx}
$^{1}$Instituto Nacional de Astrof\'isica, \'Optica y Electr\'onica, Luis Enrique Erro 1, Tonantzintla, 72849, Puebla, M\'exico\\
$^{2}$Centro de Astrobiolog\'ia (CAB, CSIC-INTA), ESAC Campus Camino Bajo del Castillo, s/n, Villanueva de la Ca\~nada, 28692, Madrid, Spain\\
$^{3}$Lund Observatory, Box 43, SE-22100 Lund, Sweden\\
}

\date{Accepted XXX. Received YYY; in original form ZZZ}

\pubyear{2021}

\begin{document}
\label{firstpage}
\pagerange{\pageref{firstpage}--\pageref{lastpage}}
\maketitle

\begin{abstract}

Planets orbiting intermediate and low-mass stars are in jeopardy as their stellar hosts evolve to white dwarfs (WDs) because the dynamics of the planetary system changes due to the increase of the planet:star mass ratio after stellar mass-loss. In order to understand how the planet multiplicity affects the dynamical stability of post-main sequence (MS) systems, we perform thousands of N-body simulations involving planetary multiplicity as the variable and with a controlled physical and orbital parameter space: equal-mass planets; the same orbital spacing between adjacent planet's pairs; and orbits with small eccentricities and inclinations. We evolve the host star from the MS to the WD phase following the system dynamics for 10 Gyr. We find that the fraction of dynamically active simulations on the WD phase for two-planet systems is $10.2^{+1.2}_{-1.0}$--$25.2^{+2.5}_{-2.2}$~$\%$ and increases to $33.6^{+2.3}_{-2.2}$--$74.1^{+3.7}_{-4.6}$~$\%$  for the six-planet systems, where the ranges cover different ranges of initial orbital separations. Our simulations show that the more planets the system has, the more systems become unstable when the star becomes a WD, regardless of the planet masses and range of separations. Additional results evince that simulations with low-mass planets (1, 10~$\mathrm{M_\oplus}$) lose at most two planets, have a large fraction of systems undergoing orbit crossing without planet losses, and are dynamically active for Gyr time-scales on the WD's cooling track. On the other hand, systems with high-mass planets (100, 1000~$\mathrm{M_\oplus}$) lose up to five planets, preferably by ejections, and become unstable in the first few hundred Myr after the formation of the WD.

\end{abstract}




\section{Introduction}
\defcitealias{maldonado2020}{Paper~I}
\defcitealias{maldonado2020b}{Paper~II}
\defcitealias{maldonado2021}{Paper~III}

 In the past few decades, observations of the circumstellar environment of a few percent of white dwarfs (WDs) have revealed that rocky material is present in the form of debris/gaseous discs (e.g. \citealt{zuckerman1987,becklin2005,gansicke2006,kilic2007,rocchetto2015,manser2016,rebassa2019,dennihy2020,melis2020}); several clumps of debris or asteroids have been observed transiting at a few WDs \citep{vanderburg2015, vanderbosch2020, vanderbosch2021b, guidry2021, farihi2022}; and major planets have also been detected \citep{gansicke2019,vanderburg2020,blackman2021}. All of this observational evidence confirms that, unexpectedly, these minor and major bodies have survived the stellar evolution of their parent stars and have reached orbital configurations close to WDs. The prevalence of this scenario is demonstrated by the observed 25-50$\%$ of WDs with metal pollution in their atmospheres \citep{zuckerman2003,koester2014}, which has originated from disrupted planets or asteroids.

The tidal forces of the expanding envelope during the Red Giant Branch (RGB) and the Asymptotic Giant Branch (AGB) of the host star lead to the engulfment of planets and minor bodies orbiting at a few au from the central star (\citealt{villaver2009};  \citealt{kunitomo2011}; \citealt{mustill2012};  \citealt{nordhaus2013}; \citealt{villaver2014};  \citealt{ronco2020}). The current theoretical paradigm to explain the phenomena close to the WD is that planets that orbit at far distances from the star and that survive the stellar evolution are responsible for scattering minor bodies, or each other (\citealt{debes2002,bonsor2011,frewen2014,mustill2018,maldonado2021}, hereafter \citetalias{maldonado2021}). This excites their orbital eccentricity and sends them on to star-grazing orbits, where the tidal forces of the WD may disrupt them and enhance the chances of debris falling on to the WD's atmosphere, resulting in metal pollution \citep{debes2012,li2021}. 

From the theoretical point of view, several studies involving N-body simulations have analyzed the dynamical evolution of planetary systems in order to explain the observed metal pollution on WDs, by using different planetary architectures through multiple phases of stellar evolution. Ultimately, the mechanism operating is the mass-loss of the host star that can trigger dynamical instabilities that somehow may contribute to WD pollution, i.e. systems stable on the main sequence (MS) become unstable in the WD phase of the stellar host \citep{debes2002}. The exploration of one-planet \citep{bonsor2011,debes2012,frewen2014,veras2021}, two-planet \citep{smallwood2018} and three-planet systems \citep{mustill2018}, interacting with planetesimal belts similar to the Edgeworth-Kuiper belt and the Asteroid Belt in the Solar System, partially explains the accretion rates observed in polluted WDs, when some stringent conditions are fullfilled, such as a planetesimal disc orders of magnitude more massive than the Asteroid Belt of the Solar System \citep{bonsor2011,debes2012} or low-mass planets with very eccentric orbits (e $>$ 0.4, \citealt{frewen2014},\citealt{mustill2018},\citealt{veras2021}).

Furthermore, other works studying the stability of planetary systems through different stellar evolution phases with multiple planetary configurations such as two planets (\citealt{veras2013,veras2013b,voyatzis2013,veras2018,ronco2020}); three  planets (\citealt{mustill2014,mustill2018}) and four and more planets \citep{veras2015,veras2016b,zinc2020} have connected the rate of dynamical instabilities with WD pollution.  Particularly, in \citet{maldonado2020,maldonado2020b}, hereafter \citetalias{maldonado2020,maldonado2020b}, and in \citetalias{maldonado2021}, we used planetary systems observed orbiting main sequence stars, rescaled to larger orbital radii so that they survive stellar evolution. We studied the rates at which these systems experience instabilites, which can include not only collisions or ejections of planets, but also orbit crossing or even weak scattering among the planets (since the semimajor axis changes thus induced can still destabilise any asteroids and comets in the system). We find that, considering all types of instability, 
two- and three-planet systems can not explain the observed prevalence of polluted WDs. On the other hand, systems with a high planetary multiplicity (i.e. with four, five and six planets) have a rate of unstable systems comparable to the observed fraction of polluted WDs. Hence, the more planets a system has, the more likely it is to have a dynamical instability that may contribute to WD pollution.

 The planetary architectures simulated in \citetalias{maldonado2020,maldonado2020b,maldonado2021} have a mixture of planet masses, eccentricities and planet separations since they were based on real systems. However, the multiplicity of planets is only one of the parameters that may determine the rate of instabilities. Disentangling the physical and orbital parameter space in multiple-planet systems is needed in order to prove that the planetary multiplicity is indeed playing the major role in producing planet--planet scattering events on the WD phase. To accomplish this objective, in this work we build new templates of planetary systems with multiple planets using a more controlled parameter space, in which they have the same planet masses and planet separations, with similar eccentricities and inclinations and the only variable is the number of planets in the systems.

In Section 2 we describe how we build the new planetary templates that we evolve from the MS to the WD phase of the host star, in Section 3 we present the results, in Section 4 we discuss them and finally in Section 5, we present the conclusions of this study.

\section{Numerical simulations set-up}
\label{setup}
 
In this work, we simulate planetary systems with two, three, four, five and six planets  with four different sets of equal masses. To solve the dynamics of the systems, we use the N-body package $\textsc{MERCURY}$ \citep{chambers1999}, which has been updated by \citet{veras2013b,mustill2018} to take into account the changes of stellar mass and radius throughout the stellar evolution phases. For this study, we use the RADAU integrator and a tolerance parameter of 10$^{-11}$ as in \cite{mustill2018}; \citetalias{maldonado2020,maldonado2020b,maldonado2021}. The integration time for each simulation is set to 10 Gyr and the output is given with an interval of 1 Myr. Planets that collide with each other, or with the stellar envelope, or those that are ejected by crossing the standard Hill ellipsoid with a size of $\sim$ $10^5$ au in the solar neighborhood \citep{veras2013c,veras2014c}, are removed from the integration. If a planet collides with the stellar host, the mass of the star is increased accordingly.  

As in our previous work, we choose the stellar evolution model better suited for the purposes of this study, a 3 $\mathrm{M_\odot}$ host star. This is motivated by the fact that polluted WDs have a mean mass of $\sim$ 0.7 $\mathrm{M_\odot}$ \citep{koester2014} which corresponds to the MS mass adopted according to a standard initial-to-final mass relation (e.g. \citealt{kalirai2008}). The stellar evolution model is produced by the SSE code \citep{hurley2000}, adopting a Reimers mass-loss parameter $\eta=$ 0.5 and solar metallicity. The timescales of the different evolution phases of the  3$\mathrm{M_\odot}$ star are: up until $\sim$ 377.5 Myr for the MS; between 377.5 and 477.6 Myr for the RGB and AGB phases, where it losses 75 $\%$ of its mass; and after 477.6 Myr, when the WD phase starts. Most likely due to observational biases, a typical stellar host in the observed multiple-planet systems have a mean mass $\sim$ 1 $\mathrm{M_\odot}$ (exoplanet.eu, \citealt{schneider2011}). Considering a star of such mass, it would take $\sim$ 10 Gyr to evolve from the MS to the WD phase, and it would be unfeasible to perform thousands of simulations in a reasonable computational time. Moreover, most Solar-mass stars have not had time to evolve to become WDs within the age of the Universe.

\subsection{Planet masses}
In this work, we test four different planet masses: 1, 10, 100, and 1000 $\mathrm{M_{\oplus}}$. All planets have the same mass in each multiple-planet template. When the radius is needed we calculate  it using the mass--radius relation proposed by \citet{chen2017}, as in \citetalias{maldonado2020,maldonado2020b,maldonado2021}. To give an overview on how the parameters used in this work compare with published studies, in the upper panel of Fig. \ref{inidel} we show the planet mass as a function of the number of planets, where the green squares represent the planet masses tested in this work. The values adopted in previous works in the literature are shown with different symbols (see the legend and the references given in the figure's caption). The box plots correspond to the mass distributions used in \citetalias{maldonado2020,maldonado2020b,maldonado2021}. The medians of each box are displayed as gray lines inside the boxes,  the upper whisker extents from the third quartile Q3 to the lower datum below Q3+1.5(Q3-Q1), with Q1 the first quartile and the lower whisker goes from Q1 to the lowest datum above Q1-1.5(Q3-Q1) in each box.  
 The outliers are shown with the small gray dots. The gray triangles in the upper part of each box distribution depict that there are mass values larger than 5000 $\mathrm{M_\oplus}$ in the samples.

\subsection{Planet spacing}
\label{delta}
\begin{figure}
\begin{center}
\begin{tabular}{cc}
\includegraphics[width=9cm, height=17cm]{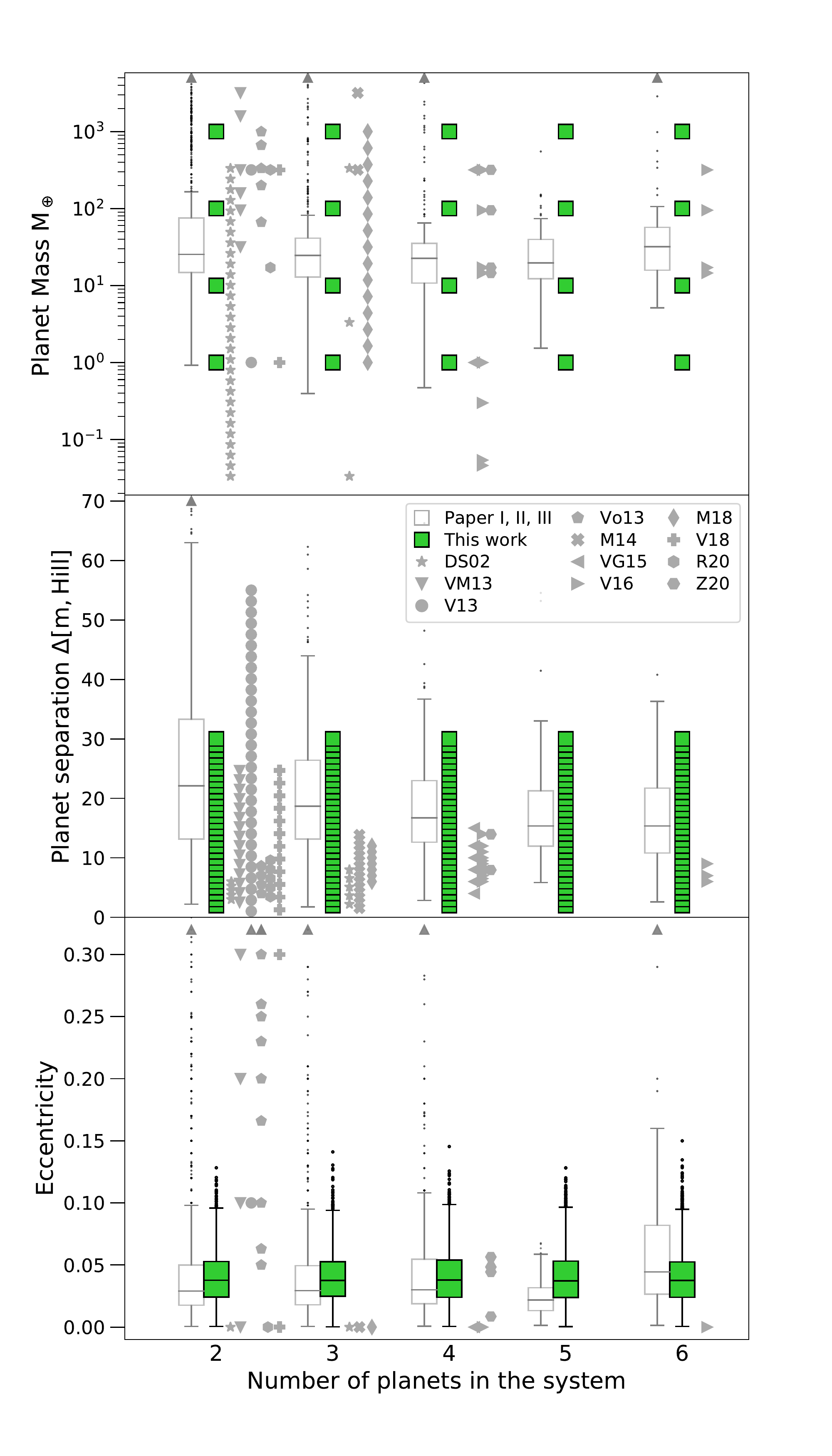}
\end{tabular}
\caption{Upper panel: Planet masses (1, 10, 100 and 1000 $\mathrm{M_\oplus}$) used in this work are shown as green squares.   Middle panel: Planet separations $\Delta$ in mutual Hill radius units used in this work are displayed as green squares, covering the range from 2 to 30.  Lower panel: Eccentricity distributions used in this work are marked as green box-plots. The parameter space explored in  \citetalias{maldonado2020,maldonado2020b,maldonado2021} appears as white box-plots. The gray triangles in the top of the figure indicates the existence of larger values than the maximum value in the Y-axis. The gray symbols appearing in the right-side of the green symbols refer to the parameters used in previous studies according to the number of planets simulated. The acronyms on the legend's figure: DS02--\citet{debes2002}, VM13--\citet{veras2013}, V13--\citet{veras2013b}, Vo13--\citet{voyatzis2013}, M14--\citet{mustill2014}, VG15--\citet{veras2015}, V16--\citet{veras2016b}, M18--\citet{mustill2018}, V18--\citet{veras2018}, R20--\citet{ronco2020}, Z20--\citet{zinc2020}. } 
\label{inidel}
\end{center} 
\end{figure}

In our simulations, planets are separated keeping equal distances (measured in mutual Hill radii) between adjacent pairs. The planet separation in terms of mutual Hill radius units is defined as $\Delta=(a_j-a_i )/\mathrm{R_{m,Hill}}$, where $a_i$, $a_j$ are the semimajor axes in planets $i$, $j$, with $a_j > a_i$, and the mutual Hill radius is defined as:
\begin{equation}
\mathrm{R_{m, Hill}}=\Bigl(\frac{a_j+a_i}{2}\Bigr)\Bigl(\frac{m_i+m_j}{3M_*}\Bigr)^{1/3},    
\end{equation}
where $m_i$, $m_j$ are the planet masses and $M_*$ is the mass of the central star. In this work we test 29 $\Delta$ values, ranging from 2 to 30 in steps of 1 mutual Hill radius. In the  middle panel of Fig. \ref{inidel}, we display the $\Delta$ values explored (the green squares) as a function of the number of planets for the smallest planet mass. $\Delta$ values used in previous works are shown as grey symbols. References are provided in the figure caption. As in the upper panel, the box plot show the $\Delta$ distributions used by us in \citetalias{maldonado2020,maldonado2020b,maldonado2021}.

In each planetary system, we place the innermost planet at $a_0$= 10 au. This distance is chosen in order to avoid the influence of tidal forces on the planets and to ensure the survival \textrm{of the planet} to engulfment on the AGB phase of evolution of the host star \citep{villaver2009,mustill2012,villaver2014}. The consecutive planets are placed at distances where $\Delta$ is kept the same between the adjacent planet pairs  in the range between 2 and 30 mutual Hill radii. Nevertheless, in systems with four, five and six  planets with planets of 1000 $\mathrm{M_\oplus}$, the outermost planet for some of the largest $\Delta$ values can reach a larger distance than the semimajor axis at which  we have set up $\textsc{MERCURY}$ to
consider a planet ejection. For this reason,  in the following and for the statistics
the maximum $\Delta$ that we consider in the four-, five- and six-planet systems with 1000 $\mathrm{M_\oplus}$ planets is 21, 19 and 17, respectively. 

Furthermore, we would like to mention that there is a maximum planet separation $\Delta_{\mathrm{max}}$, since the semimajor axis $a_j$ $\rightarrow$ $\infty$ as $\Delta_{\mathrm{ max}}\rightarrow 2(\frac{m_i+m_j}{3M_*})^{-1/3}$ \citep[cf.][]{mustill2014}. For planetary systems involving 1, 10, 100 $\mathrm{M_\oplus}$, testing $\Delta\leq 30$ is completely feasible. However, for 1000 $\mathrm{M_\oplus}$ planets, $\Delta\to\Delta_{\mathrm{ max}}= 22.88$, which means that for these simulations, the maximum $\Delta$ we can test is 22.

\subsection{Orbital eccentricities}
Besides the mass and planet separation $\Delta$, eccentricities, and orbital inclinations are also required for the initial set-up of the simulations. \citet{moorhead2011,vaneylen2015,pu2015} have demonstrated that the eccentricity and inclination in observed multiple-planet systems are well described by Rayleigh distributions. Thus, in this work, we select randomly for each planet the eccentricity and inclination from a Rayleigh distribution with parameters $\sigma=0.03$ and $\sigma=1.12^\circ$ \citep{xie2016}.  In the lower panel of Fig. \ref{inidel}, we show  the eccentricity distributions used in this work as green box plots. The outliers correspond to eccentricities in the tail of the Rayleigh distribution (values larger than 0.15 were not generated in the random selection from the Rayleigh distribution). For comparison, eccentricity values from previous studies are also displayed with gray symbols (references are given in the figure's caption). We can see that previous two-planet studies have performed simulations with several eccentricity values, but works that simulated planetary systems with three and more planets have mainly tested the circular case. The only exception is our \citetalias{maldonado2020,maldonado2020b,maldonado2021} where we simulated a distribution of eccentricities based on parameters obtained from observations of planetary systems (the white box symbols). We see a slight difference between the white and green box plots: it is due to the fact that the eccentricities in the white box plots were done using a $\sigma=0.02$ \citep{pu2015} for  transiting planets and in this work both eccentricity and inclination $\sigma$ parameters are taken from the mean values derived from the $\it{Kepler}$ multiple-planet sample analyzed by \citet{xie2016}.

We have randomly selected  the angular phases in the planet's orbit, i.e. argument of the pericentre, mean anomaly and ascending node, from uniform distributions from 0 to 360$^\circ$.

\section{Results}
The sample of simulations that we use in this work is built as follows: 
 for each planet multiplicity of two, three, four, five, and six planets, we build 4 sets of systems with equal mass planets considering 1, 10, 100, 1000 $\mathrm{M_\oplus}$ masses; for each set of planet mass, we use 29 planet separations in $\Delta$  from 2 to 30 mutual Hill radii (with the exception in planetary systems with four, five, and six planets involving 1000 $\mathrm{M_{\oplus}}$  in which we consider $\Delta$ in the range of 2--21, 2--19 and 2--17, respectively); for a single $\Delta$ value, we make 10 runs per planetary configuration, keeping constant the planet mass and $\Delta$, varying randomly the eccentricity, inclination and the three phase angles, accordingly. Therefore, we simulate 5310 planetary systems in total, where 1080 simulations are from two- and three-planet systems, 1070 from four-planet systems, and 1050 and 1030 from five- and six-planet systems, respectively.

In total, we perform 5310 $N$-body simulations of planetary systems with two to six planets, evolving a 3 $\mathrm{M_\odot}$ host from the MS to the WD phase. We track the incidence of various instabilities in the systems. The instabilities include not only ejection of planets and collisions between planets or planets and the star, but also orbit crossing without loss of a planet, and weak scattering without orbit crossing but involving a moderate change in a planet's semimajor axis -- in short, anything that may destabilise asteroids in the system.

\begin{figure}
\begin{center}
\begin{tabular}{c}
\includegraphics[width=8.6cm, height=16cm]{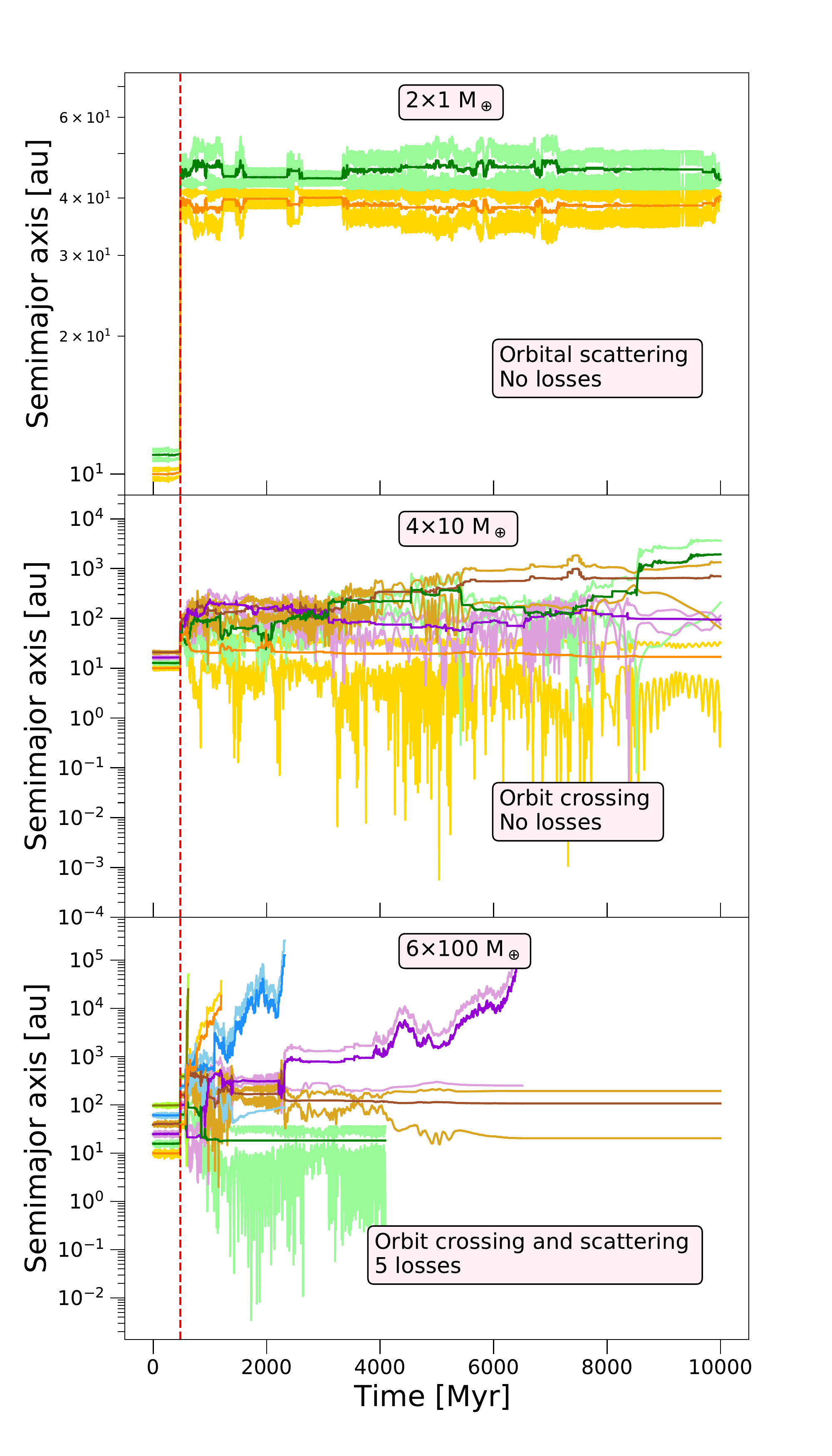}
\end{tabular}
\caption{10 Gyr of semimajor axis evolution of multiple-planet systems dynamically active on the WD phase. The solid lines in different colours refer to the semimajor axis of different planets. The light versions of each colour depict the apocentre and pericentre of each planet. The red vertical dashed lines shows the beginning of the WD phase. In the upper panel we show a two-planet system with orbital scattering in the planet's semimajor axis evolution where none of the planets are lost. The planets have a mass of 1 $\mathrm{M_\oplus}$. In the middle panel, we show a four-planet system with orbit crossing among the planet's orbits and no planet is lost the entire simulated time. Planets have a mass of 10 $\mathrm{M_\oplus}$. In the lower panel we present a six-planet system where 5 planets are lost (4 ejections and 1 planet-star collision [green]) and only one planet remains. It is a system with 100 $\mathrm{M_\oplus}$ planets. }
\label{exam}
\end{center} 
\end{figure}

In Fig. \ref{exam} we show, as illustrative examples from our simulations, the semimajor axis evolution of three planetary systems that experience different dynamical instabilities on the WD phase. The planets' semimajor axes are displayed in solid lines with different colours and the lighter version of each color shows the pericentre and apocentre of each planet. The red vertical dashed line marks the time when the 3 $\mathrm{M_\odot}$ stellar host becomes a WD (477.6 Myr). 

The upper panel of Fig. \ref{exam} shows a two-planet system with 1 $\mathrm{M_\oplus}$ planets and planet separation $\Delta=11$. We see that both planets survive the 10 Gyr of evolution; however, their semimajor axes show a variation that we called orbital scattering. We define that a planet experiences orbital scattering when its semimajor axis changes more than 5 $\%$ with respect to the semimajor axis at the beginning of the WD phase (i.e. $a(t)/a(t=478 \mathrm{Myr})>5\%$).  

The middle panel of Fig. \ref{exam} represents a four-planet system with  10 $\mathrm{M_\oplus}$ planets and $\Delta=$ 13. In this case, all the planets survive the entire simulated time, despite experiencing an instability very soon after mass-loss that leads to orbit crossing and ongoing scattering over the whole integration period (it is a Hill unstable system). We also see that the eccentricity of some planets, specially in planet 1 (yellow), is highly excited,
causing the planet's pericentre to reach distances of $\sim1\mathrm{\,R}_\odot$ from the WD at 3 different times: 5039, 5236 and 7313 Myr.

Finally, the lower panel of Fig. \ref{exam} depicts a six-planet system involving 100 $\mathrm{M_\oplus}$ planets separated by $\Delta=$ 11. This system is Hill and Lagrange unstable: it shows orbit crossing among the planets, and five planets are lost, four by ejections (planet 1, 3, 5 and 6) and one by collision with the star (planet 2, green colour). Before planet 2 is lost, it experiences several episodes of small pericentre distance that may induce tidal circularization. These multiple-planet systems are good examples of systems with dynamical instabilities that may contribute to WD metal pollution, that could be even enhanced if putative asteroid belts are present.     

\subsection{Instabilities on the MS}
\label{msuns}

There is a fraction of simulations where orbit crossing and/or planet losses happen during the MS phase of the host star. In numbers, we find that 172/1080 (15.9$\%$) in the two-planet, 344/1080 (31.9$\%$) in the three-planet, 367/1070 (34.3$\%$) in the four-planet, 394/1050 (37.5$\%$) in the five-planet and 416/1030 (40.4$\%$) in the six-planet simulations are unstable on the MS. The early instabilities occur because a fraction of planet pairs in these systems are separated by distances where mean motion resonances (MMR) overlap. Two planets are in MMR when the ratio of orbital periods between the planets is expressed as the ratio of two integers $(p+q)/p$, with $p$ the integer and $q$ the order of the resonance. 

We estimate the planet separation of the chaotic zone by using $(a_j-a_i)/a_i=1.8e^{1/5}(m_p/M_*)^{1/5}$, with $m_p=m_i=m_j$ the planet mass and $e$ the planet eccentricity (c.f. \citealt{mustill2012b}).  Despite the fact that the  latter equation predicts better the stability of systems with two equal-mass planets, it give us clues where systems with a higher number of planets are prone to have early instabilities when the adjacent planet pairs are separated less than such limiting distance. By using the eccentricity values randomly selected from the Rayleigh distribution in this work, the average of the limiting distance $\Delta$ in terms of mutual Hill radius for 1, 10, 100 and 1000 $\mathrm{M_\oplus}$ is respectively  6.5, 4.7, 3.4, 2.4, implying that simulations with $\Delta$ lower than the limiting distances previously calculated imminently lose planets or have orbit crossing on the MS. The rest of unstable systems in the MS have adjacent pairs of planets whose semimajor axis ratio is found within the 5$\%$ from the corresponding semimajor axis ratio in some first and second order MMR, mainly the 3:1, 2:1, 5:3, 3:2, 4:3, 5:4, 6:5, influencing the dynamics of the system (with the exception of 12 simulations in systems of four, five and six planets with 1000 $\mathrm{M_\oplus}$ having adjacent planet pairs within the 8~$\%$ of the 2:1 resonance).  

Due to the fact that the resonant systems need a specific orbital configuration in order to be dynamically stable and that the finding of such configuration is beyond the scope of this work, we removed the systems in MMR with unstable outcomes in the MS for the final statistics. However, we do consider for the statistics the resonant systems in which the random selection of the orbital angles prevents the early instability on the MS.

\subsection{Non-adiabatic mass-loss regime}
\label{nonadiab}

The adiabatic approximation is defined as the condition fulfilled when the stellar mass-loss time-scale is much larger than the orbital time-scale of a planet. In this case, the semimajor axis evolution is given by
\begin{equation}
\Bigl(\frac{da}{dt}\Bigr)_{\mathrm{ad}} = -\frac{a}{\mu}\frac{d\mu}{dt},
\end{equation}
with $\mu$ the planet--star mass ratio. In this regime, the eccentricity remains almost unchanged. 

The effects of adiabatic and non-adiabatic mass-loss on the dynamical evolution of single planets have been explored  by e.g. \citealt{veras2011}.  In general, they find that planets located at orbital distances $\geq$ 1000 au with different eccentricities are likely to suffer an ejection from the system during the mass-loss of the star: the planet's eccentricity will increase drastically, and as a consequence the orbit of the planet becomes hyperbolic and the planet is imminently ejected during or just after the major mass-loss event.

The non-adiabatic regime produces planet ejections due to the excitement of the planet's eccentricity and we consider that it is important to know when the non-adiabatic effects become important in our simulations. Therefore, we perform additional simulations of single planets orbiting a 3$\mathrm{M_\odot}$ and evolving the systems for 500 Myr (from the MS to a few Myr after the beginning of the WD phase). We test one-planet systems with the four planet masses studied in this work (1, 10, 100 and 1000 $\mathrm{M_\oplus}$). The eccentricity, inclination and the phase angles in the planet's orbit are randomly selected from the Rayleigh and uniform distributions described in Section \ref{setup} for each template. 

In the first trial, we test different initial semimajor axis ($a_\mathrm{initial}$) values, from 50 to 1000 au in steps of 50 au and we find that the eccentricity increases as the planet is located at further distances from the host star; we do not find any difference in the eccentricity change among the simulations with different planet masses. We find that around 250 au the final eccentricity ($e_\mathrm{final}$) increases to $\sim$ 0.2. 

\begin{figure}
\begin{center}
\begin{tabular}{c}
\includegraphics[width=8.5cm, height=6.5cm]{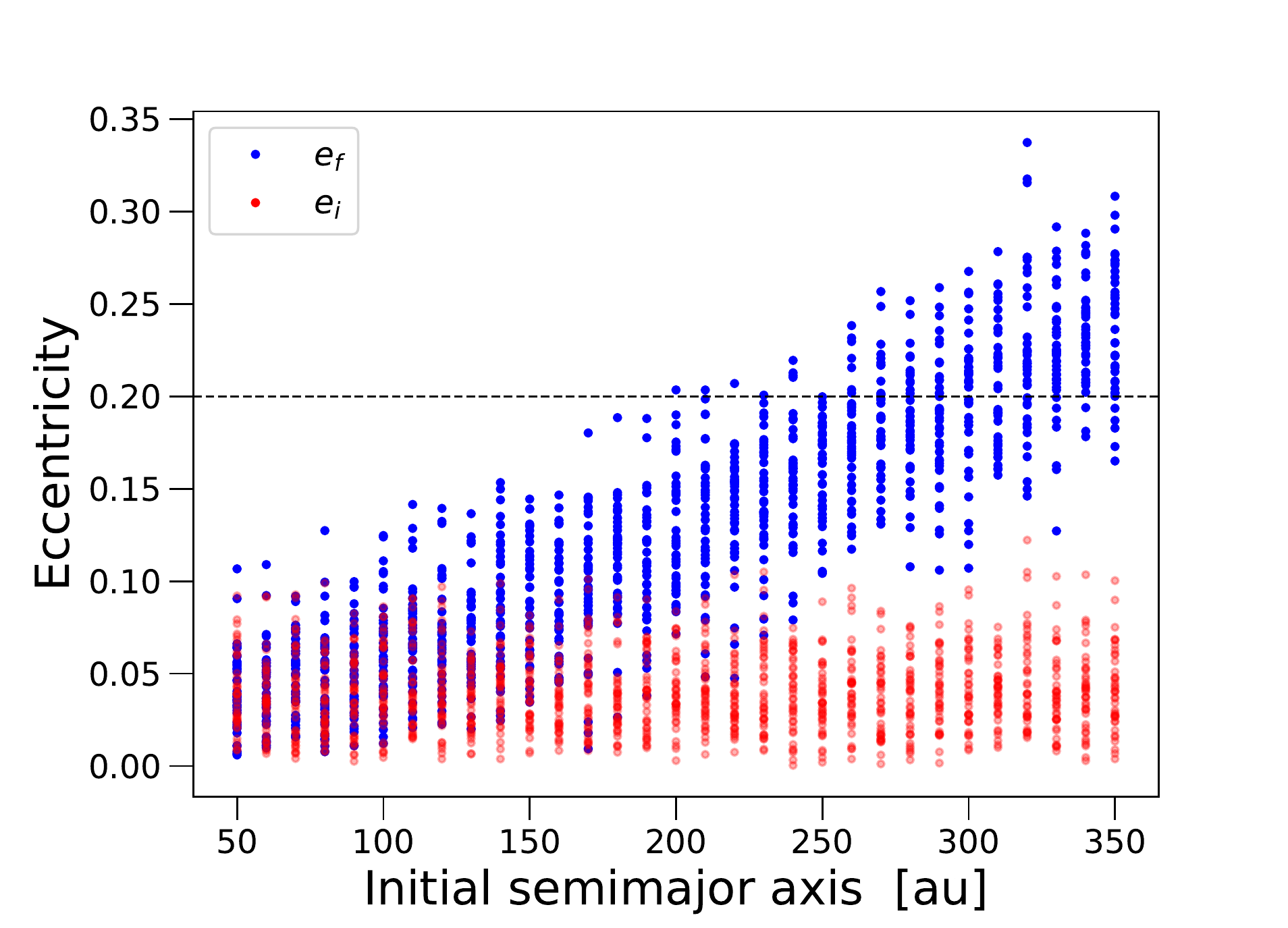}
\end{tabular}
\caption{Eccentricity as a function of initial semimajor axis in one-planet systems. The red dots refer to eccentricities at the beginning of the integration and the blue dots show the final eccentricity taken just after the WD forms (478 Myr for a 3 $\mathrm{M_\odot}$ host). The horizontal black dashed line marks $e=0.2$. The semimajor axis step is 10 au, and 40 simulations were performed in each step (10 simulations per planet mass, varying the eccentricity and inclinations from the Rayleigh distributions  and phase angles from the uniform distributions defined in section \ref{setup}, respectively, in each simulation). }
\label{onepl}
\end{center} 
\end{figure}

In the second trial, we make additional simulations in order to refine $a_\mathrm{initial}$ from  50 to 350 au, with steps of 10 au. This time, we test the four planet masses with the orbital parameters randomly selected in each run and we make 10 runs per planet mass and per $a_\mathrm{initial}$ value.  In Fig.\ref{onepl}, we show eccentricity as a function of the initial semimajor axis in each one-planet simulation. Due to the fact that we do not find any difference among the eccentricities for different planet masses, we display in the blue dots the final eccentricities in each system  taken after the formation of the WD (478 Myr). For comparison, we also show with the red dots the initial eccentricities randomly selected from the Rayleigh distribution in each simulation. The horizontal black dashed line depicts $e_\mathrm{final}=0.2$ for better visualization along the x-axis. We see that the eccentricity is increasing as the planet is located further from the central star and for $a_\mathrm{initial}< 200$ au, $e_\mathrm{final}<0.2$.  Moreover, we see that the initial and final eccentricities converge as $a_\mathrm{initial}$ goes to 50 au, which it is expected in the adiabatic regime.

In this work, the planetary templates with large $\Delta$ values have planets at orbital distances where non-adiabatic ejections may affect the dynamical evolution. Indeed, we  find for planetary systems with 100 $\mathrm{M_\oplus}$ planets, 27 simulations in the five-planet case ($\Delta\geq$ 28) and 70 simulations in the six-planet case ($\Delta\geq$ 24) where planets are ejected or have hyperbolic orbits when the star becomes a WD (at 478 Myr). Moreover, in systems with 1000 $\mathrm{M_\oplus}$ planets, we have 30 simulations in the three-planet case ($\Delta\geq$ 20), 58 simulations in the four-planet case ($\Delta\geq$ 16), 67 simulations in the five-planet case ($\Delta\geq$ 13) and 68 simulations in the six-planet case ($\Delta\geq$ 11) where non-adiabatic ejections happen. Since the ejections in all these simulations are produced by non-adiabatic effects and not by dynamical interactions among the planets, we decide to remove these simulations for the final statistics.

\subsection{Dynamical instabilities on the WD phase }
\label{wduns}
We take into account simulations dynamically active on the WD phase (having planet losses, orbit crossing and orbital scattering in at least one planet, as the examples depicted in Fig. \ref{exam}). As described in sections \ref{msuns} and \ref{nonadiab}, for the final statistics of unstable systems we remove simulations where planet losses and orbit crossing happens on the MS. We do not consider simulations where at least one planet is ejected by non-adiabatic mass-loss.

 Due to the fact that many observed protoplanetary disks have extensions $\lesssim$ 200 au (e.g. \citealt{tazzari2017,garufi2020,sean2020}) and based on the previous results from the one-planet tests performed in section \ref{nonadiab}, we check in our multiple-planet templates with the highest number of planets which planet separation $\Delta$ places all planets within 200 au in each system. For six-planet systems with 1 and 10 $\mathrm{M_\oplus}$, the outermost planet is always located at $a< 200$ au with $\Delta\leq 30$. However, for 100 and 1000 $\mathrm{M_\oplus}$, the sixth planet fulfils the condition only with $\Delta\leq 14$ and $6$, respectively. We choose $\Delta = 14$ as one of the threshold limits in the planet separation space in order to analyze our results. We make this choice because for $\Delta =6 $, we would have very few or no simulations with dynamical instabilities on the WD phase to count after removing the MS unstable simulations for the statistics.

 We consider $\Delta=$ 17 and 30 as additional $\Delta$ thresholds in order to analyze the fraction of unstable simulations. The former considers all the planet masses tested in this work with the largest planet separation simulated ($\Delta$=17 in the six-planet systems with 1000 $\mathrm{M_\oplus}$ planets). The latter considers only planetary systems with 1, 10 and 100 $\mathrm{M_\oplus}$ planets since $\Delta=30$ is the largest planet separation considered in this study and is only tested for those planet masses ($\Delta_{\mathrm{ max}} = 22.88$ for 1000 $\mathrm{M_\oplus}$ planets; see section \ref{delta}).

\begin{figure*}
\begin{center}
\begin{tabular}{c}
\includegraphics[width=18cm, height=9.5cm]{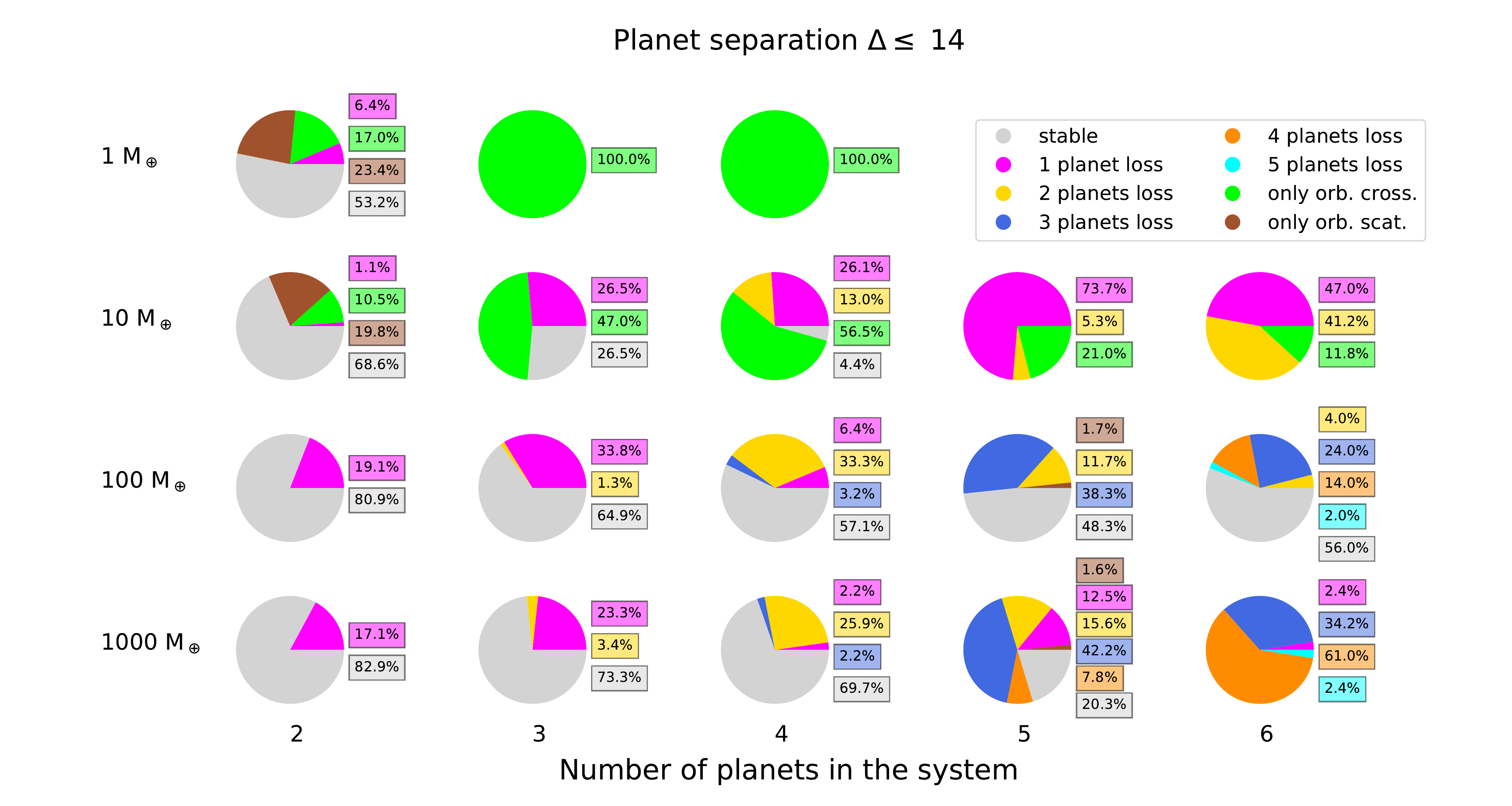}
\end{tabular}
\caption{Pie charts showing fractions of simulations with different dynamical outcomes on the WD phase. Rows indicate the planet mass and columns refer to the number of planets in the system. In gray colour we refer to simulations that remain completely stable during the 10 Gyr of simulated time. The green colour refers to simulations where no planets are lost but do have at least one orbit crossing. The brown colour depicts simulations where there are neither planet losses nor orbit crossing but at least one of the planets have orbital scattering. The planet losses are marked with colors as written in the figure's legend. The percentage of unstable simulations are highlighted in the squares on the right side of each pie chart for better visualization.  Simulations with $\Delta\leq$ 14 are considered here.  }
\label{pie14}
\end{center} 
\end{figure*}

\begin{figure*}
\begin{center}
\begin{tabular}{c}
\includegraphics[width=18cm, height=9.5cm]{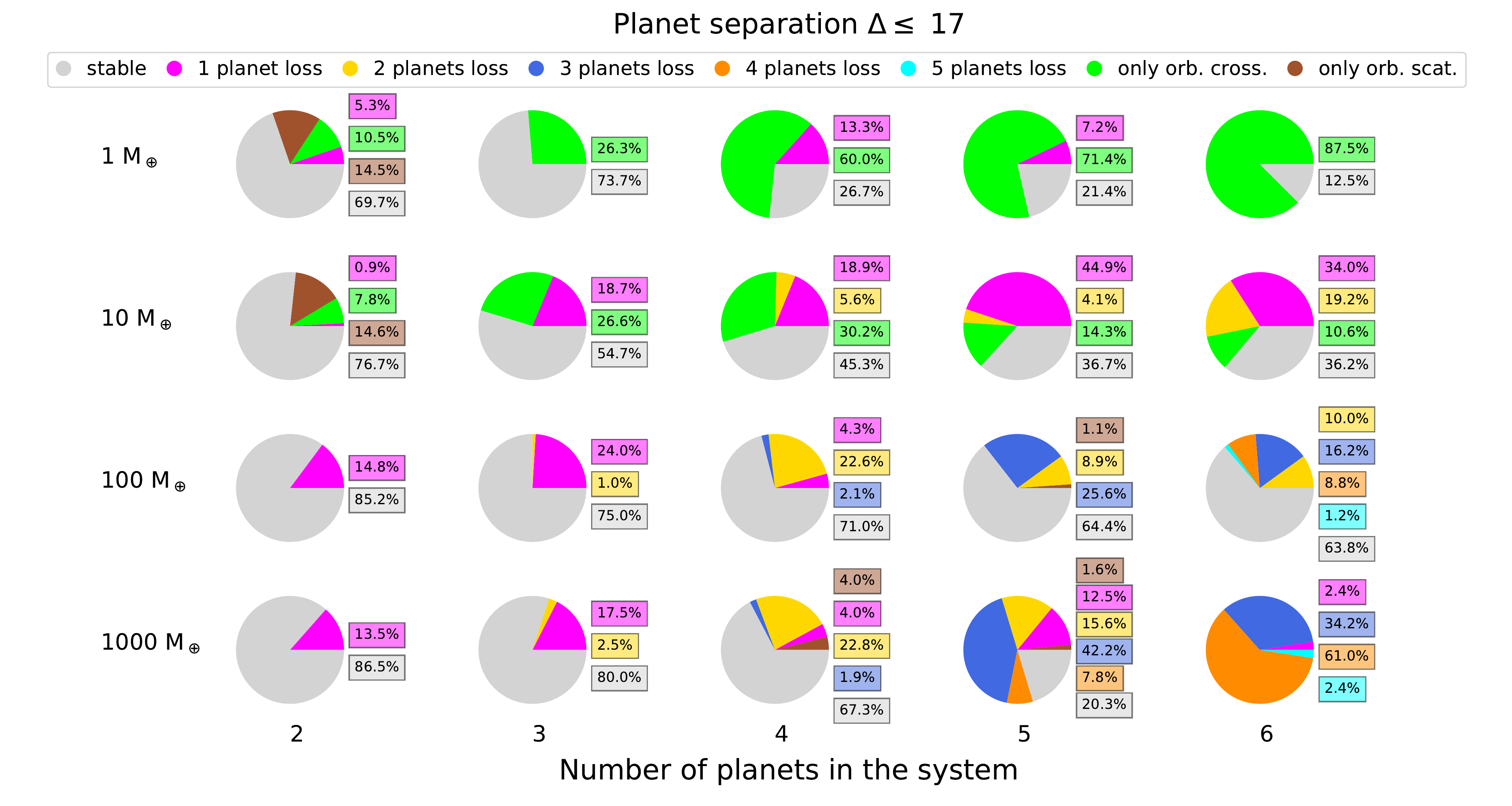}
\end{tabular}
\caption{The same as in Fig. \ref{pie14} but with planetary systems having $\Delta\leq$ 17.  }
\label{pie17}
\end{center} 
\end{figure*}

\begin{figure*}
\begin{center}
\begin{tabular}{c}
\includegraphics[width=18cm, height=9.5cm]{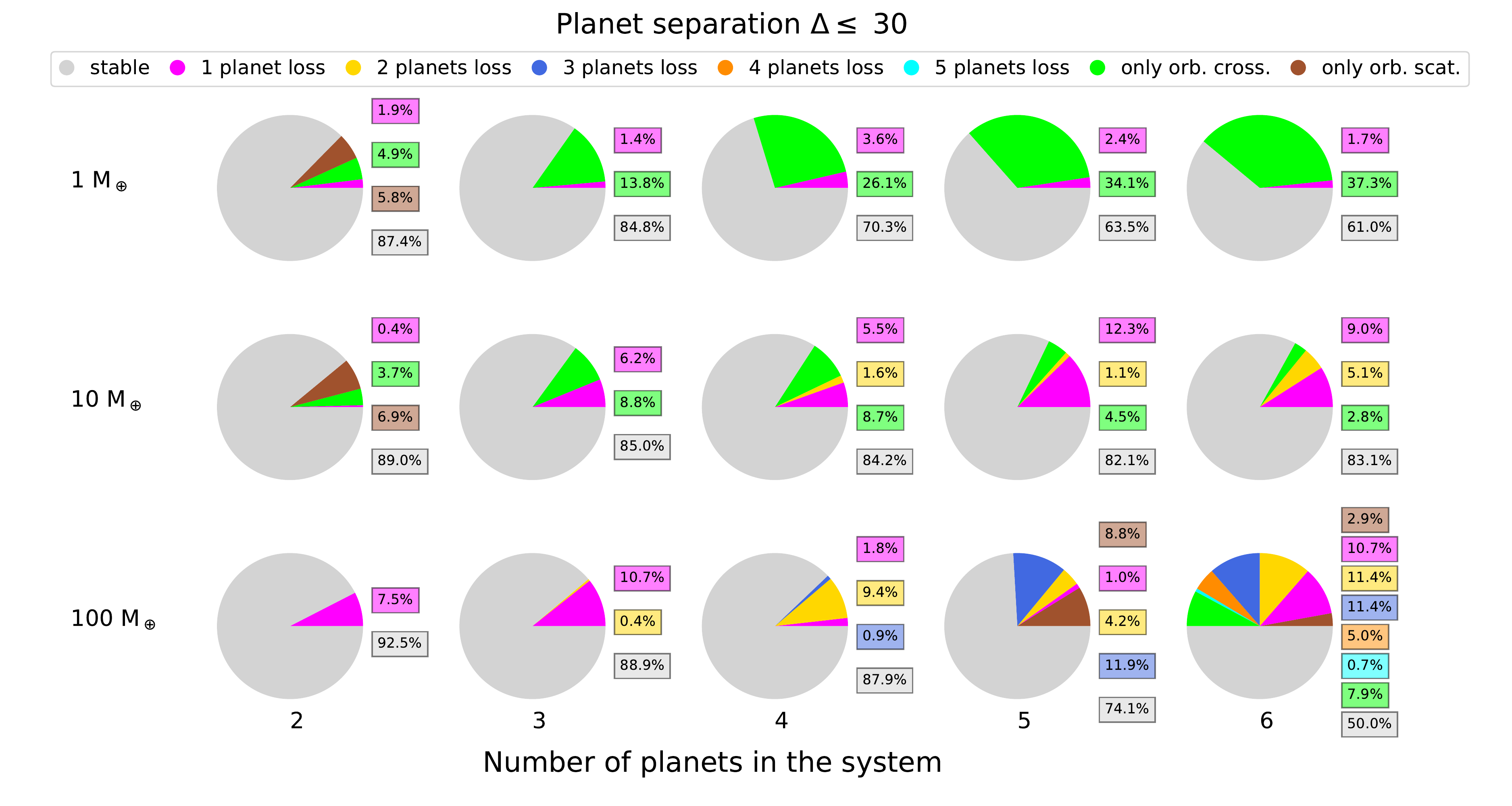}
\end{tabular}
\caption{ The same as in Fig. \ref{pie14} but with planetary systems having $\Delta\leq$ 30.}
\label{pie30}
\end{center} 
\end{figure*}

We show in Fig. \ref{pie14}, \ref{pie17} and \ref{pie30} several pie charts with the fraction of stable and unstable simulations considered for the statistics ordered in rows (each one per planet mass) and columns (each one referring to the number of planets in the system). The different colours depict different dynamical outcomes. In green, we mark simulations that do not lose any planet in the 10 Gyr but show one or more orbit crossings of the planets  (e.g. see as an example of such behaviour the middle panel in Fig. \ref{exam}). In brown, we show simulations where at least one planet have orbital scattering in the semimajor axis without orbit crossing and no planets are lost in the entire simulated time (e.g. the  upper panel in Fig. \ref{exam}). Regarding planet losses, either by Hill or Lagrange instabilities, there are cases where multiple planets are lost in an individual simulation (e.g. the  lower panel in Fig. \ref{exam}), specially in systems with five and six planets. Therefore, the pink, yellow, blue, orange and light-blue colours refer to simulations that loss 1, 2, 3, 4, 5, planets respectively. Lastly, the gray colour marks completely stable systems that only show the adiabatic expansion of the planets' semimajor axes due to the stellar mass loss when the star becomes a WD.  

We see different dynamical behaviours regarding the planet mass, the planet separation, and the number of planets considered in the systems. In Fig. \ref{pie14}, for $\Delta\leq$ 14, we first see that there are no pie charts for the five- and six-planet systems on the first row. This is due to the fact that all the simulations in these cases experience dynamical instabilities on the MS. Regarding orbit crossing without planet losses, we observe that this instability is happening in all but restricted to simulations with 1 and 10 $\mathrm{M_\oplus}$ planets (low-mass planets). Orbital scattering without planet losses is present in a fraction $\sim$ 20 $\%$ of the two-planet systems with low-mass planets. A much lesser fraction is obtained in the five-planet systems with 100 and 1000 $\mathrm{M_\oplus}$ (high-mass planets), in which the outermost planet shows the scattering. Moreover, simulations losing one planet occur in all but 4 cases, where the highest fractions are obtained in five- and six-planet systems with 10 $\mathrm{M_\oplus}$. Planetary systems losing two and more planets are restricted to  templates with four, five and six planets with 10, 100 and 1000 $\mathrm{M_\oplus}$. We do not have any simulation that loses all the planets during the 10 Gyr. The number of stable simulations decreases in all the cases as the number of planets increases.

Regarding simulations with $\Delta\leq$ 17, in Fig. \ref{pie17} we have all the rows and columns with pie charts. In general, we observe a similar dynamical behaviour with respect to Fig. \ref{pie14}. The fraction of stable simulations decreases as the number of planets increases. Simulations with orbit crossing without planet losses  are only present in systems with low-mass planets and we see an increasing trend in systems with 1 $\mathrm{M_\oplus}$ planets, as the number of planets increases. Besides having the same  cases with simulations showing only orbital scattering without planet losses as $\Delta\leq$ 14, we have a new small percentage  of simulations with orbital scattering and no planet losses appearing in the four-planet systems with planets of 1000 $\mathrm{M_\oplus}$, where the fourth planet has the scattering. We note that the fraction of stable and unstable simulations in systems with five and six planets having 1000 $\mathrm{M_\oplus}$ remains constant. It happens because non-adiabatic mass-loss ejections happen in  these simulations with $\Delta$ $\geq$ 13 and 11, respectively, they are removed from the statistics and  no extra simulations are counted for $\Delta\leq 17$ (see the analysis of section \ref{nonadiab} and Fig. \ref{pie14}). 

Finally, in Fig. \ref{pie30}, we display only planetary systems with planets having 1, 10 and 100 $\mathrm{M_\oplus}$ in which we perform simulations with $\Delta\leq$~30. We see that in all the cases the fraction of stable simulations is in the range between 50 $\%$ and 93 $\%$. We confirm that the fraction of simulations with orbit crossing and no planet losses increases as the number of planets increases in simulations with planets of 1 $\mathrm{M_\oplus}$. It does not happen for planetary systems with 10 $\mathrm{M_\oplus}$ planets,  instead, the fraction of planet losses increases, while the overall fraction of unstable systems increases only slightly. In addition to these low-mass cases, we have another fraction of simulations with orbit crossing and no planet losses in the six-planet systems with 100 $\mathrm{M_\oplus}$ planets. Regarding simulations with only orbital scattering without planet losses, we obtain the same cases as described in $\Delta\leq$ 14, and 17; however, for the simulations with high-mass planets, there is an increase in the fraction of five-planet systems experiencing weak scattering, and the six-planet systems have only a small fraction. All the planetary systems lose at least 1 planet and for systems with four and more planets, the higher the planet masses are, the more planets in a single planetary configuration get lost. The six-planet systems with planet mass of 100  $\mathrm{M_\oplus}$ have simulations that lose from 1 to 5 planets in a single template. 

 As mentioned in section \ref{nonadiab}, the planet's eccentricities are excited due to non-adiabatic effects and the further the planet is located from the central star, the greater the eccentricity excitation induced during the mass-loss of the stellar host, even if non-adiabatic ejections do not happen. For the case of six-planet simulations with 100 $\mathrm{M_\oplus}$ planets, we obtain that non-adiabatic ejections happen for $\Delta\geq$ 24. However, for $\Delta=$ 22 and 23, the sixth planet's eccentricity ranges from 0.72 to 0.93 and the fifth planet's eccentricity is in the range $0.29<e<0.44$. As a consequence, orbit crossing between the sixth and fifth planets is observed in the  7.9 $\%$ of simulations without planet losses observed in Fig. \ref{pie30} and starts from the beginning of the WD phase. A similar eccentricity evolution is observed in the five-planet systems with 100 $\mathrm{M_\oplus}$ for $26\leq\Delta\leq28$, where the fifth planet has $e\geq$ 0.75, not enough to produce orbit crossing with the fourth planet ($0.17< e <0.35$); however, the perturbations induced by the other planets to the fifth planet produce the 8.8 $\%$  of simulations having  orbital scattering without losing the planet observed in  the five-planet column in Fig. \ref{pie30}. Having dynamical instabilities in the outer planets of a planetary system could perturb a putative outer asteroid belt and some of its components may be launched towards the inner planetary system, where the inner planets might continue sending the material toward the WD, as suggested in \citet{bonsor2011,mustill2018,smallwood2018}.

\begin{figure*}
\begin{center}
\begin{tabular}{c}
\includegraphics[width=18cm, height=6cm]{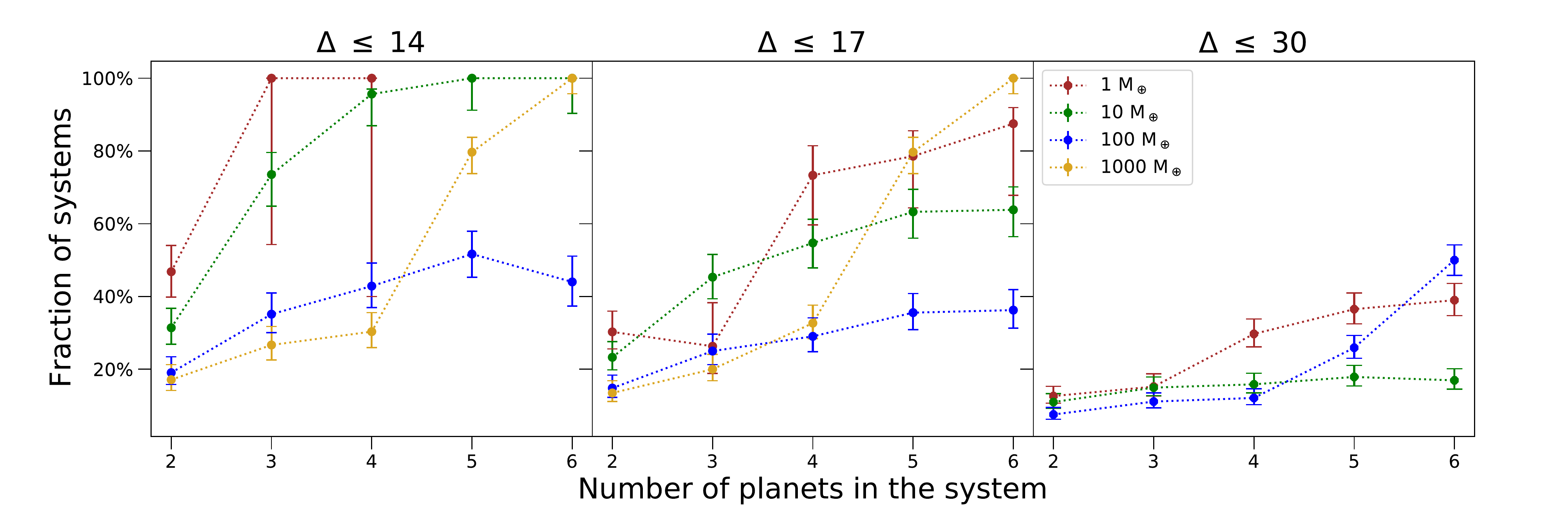}
\end{tabular}
\caption{Fraction of unstable simulations as a function of planetary multiplicity. Simulations with the same planet masses are shown in the same colour, defined in the figure's legend. In the left-hand panel we show those unstable simulations on the WD phase that may contribute to WD pollution (by having orbit crossing, orbital scattering or planet losses) whose planets have planet separations $\Delta\leq$ 14 (mutual Hill's radius units). In the middle-hand panel, simulations with $\Delta\leq$ 17 and in the right-hand panel those simulations with $\Delta\leq$ 30. In the latter case, we only plot 1, 10, and 100 $\mathrm{M_\oplus}$ simulations. 
Error bars are estimated using the binomial sampling distribution and are equivalent to 1$\sigma$ limits on a Gaussian distribution \citep{jaynes2003}.}
\label{pmass}
\end{center} 
\end{figure*}

By considering the total fraction of unstable simulations (summing the simulations that lose planets, have orbit crossing and orbital scattering), we present in Fig. \ref{pmass} three panels with the total fraction of simulations dynamically active on the WD phase with respect to the total number of simulations considered for the statistics as a function of the number of planets in the systems.  We refer to systems involving a specific planet mass by different colours (brown, green, blue and yellow for 1, 10, 100, 1000 $\mathrm{M_\oplus}$ planets, respectively). Error bars are calculated  assuming a binomial sampling distribution \citep{jaynes2003} and are equivalent  to  1$\sigma$ in Gaussian distributions. Each panel from left-hand to right-hand displays the fraction of simulations calculated  within the planet separation $\Delta\leq$ 14, 17 and 30. Globally, we can see that planetary systems with 1 $\mathrm{M_\oplus}$ are more dynamically active on the WD phase than the systems with higher mass planets, with the exception of systems with  1000  $\mathrm{M_\oplus}$ planets, where  the five- and six-planet systems with $\Delta\leq$ 14 and 17 have an increase  of unstable simulations reaching fractions higher than 70 $\%$. This increment is due to the fact that  simulations with 1000 $\mathrm{M_\oplus}$ planets are dynamically active on the WD phase with $\Delta$ values smaller than lower-mass planet simulations and non-adiabatic ejections start to appear at $\Delta$ values smaller than for the other planet masses (the $\Delta$ ranges considered for the statistics without the MS unstable simulations and after removing the non-adiabatic ejections are 6--13 and 6--11 for the five- and six-planet systems, respectively). 
The same happens to a lesser extent in five- and six-planet simulations having 100 $\mathrm{M_\oplus}$ planets for $\Delta\leq$ 30  ($\Delta$ values considered for the statistics: 8--28 and 8--23 for systems with five and six planets, respectively). Finally, we see a general tendency to have a larger fraction of unstable simulations as the number of planets increases in the systems, independently of the planet mass and the limits in $\Delta$.     

\begin{figure}
\begin{center}
\begin{tabular}{c}
\includegraphics[width=8.5cm, height=6.5cm]{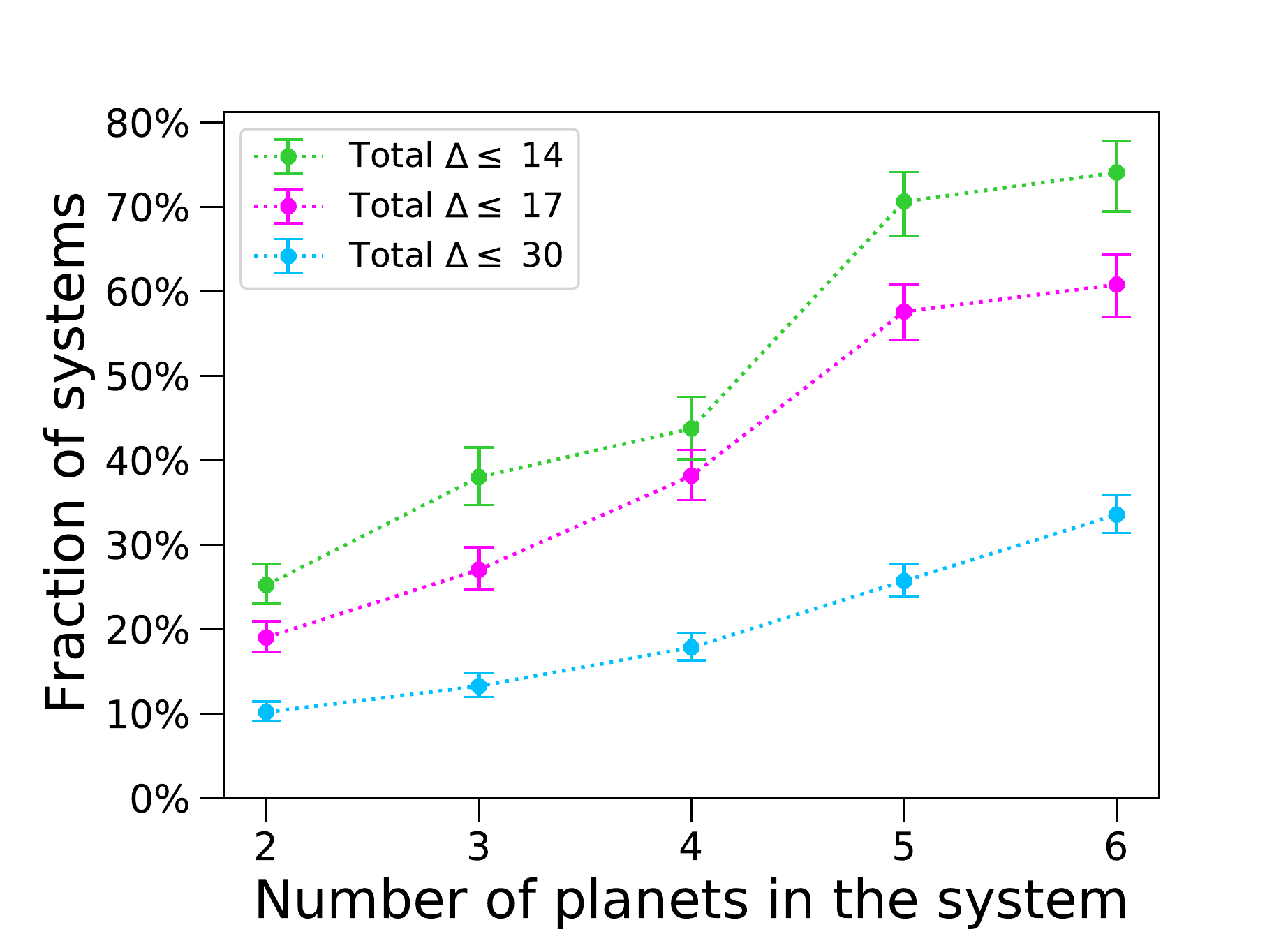}
\end{tabular}
\caption{Fraction of simulations that may contribute to WD pollution (unstable systems with respect to the total sample of simulations) as a function of the planetary multiplicity. Each color refers to calculated fractions within the thresholds in the planet separation $\Delta$,  shown in the legend's figure. Error bars are estimated as in Fig. \ref{pmass}} 
\label{totwd}
\end{center} 
\end{figure}

After adding all the unstable simulations in the four planet masses but keeping separated the $\Delta$ thresholds, in Fig. \ref{totwd} we show the total fraction of simulations dynamically active as a function of the number of planets in the system. Each colour refers to the $\Delta$ limits according to the figure's legend. The error bars are  calculated with the binomial sampling distribution \citep{jaynes2003}. We observe that higher fractions of unstable systems are obtained for simulations with $\Delta\leq$ 14 than systems with $\Delta\leq$ 17 and 30, which is an  expected result since planets separated by large mutual Hill radius are less likely to interact each other and remain completely ordered the 10 Gyr of evolution. On the other hand, we can see the clear tendency of having more dynamically active simulations as the multiplicity of planets increases for the three $\Delta$ thresholds. The fraction goes from  $10.2^{+1.2}_{-1.0}$~$\%$, $19^{+1.9}_{-1.7}$~$\%$ and  $25.2^{+2.5}_{-2.2}$~$\%$ in the two-planet systems to $33.6^{+2.3}_{-2.2}$~$\%$, $60.8^{+3.5}_{-3.8}$~$\%$ and $74.1^{+3.7}_{-4.6}$~$\%$ in the six-planet systems, for $\Delta\leq$ 30, 17 and 14, respectively.

\section{Discussion}

\subsection{Dynamical behaviour in the parameter space}

We find that the fraction of dynamically active simulations on the WD phase becomes larger as the number of planets within the planetary system increases. This result has been probed in a sample of multiple-planet systems built ad-hoc so the $\Delta$ thresholds and planetary masses and eccentricities are the same among the multiple systems tested. This result confirms the main conclusion from \citetalias{maldonado2021}, in which scaled versions of observed multiple-planet systems were simulated but in which multiplicity was just one of the parameters that could be hidden among a mix of planet masses, eccentricities and planet separations of the simulated systems. We find that the fraction of instabilities increases with the multiplicity of the system independently of the considered planet separation $\Delta$ thresholds (see Fig. \ref{totwd}) and  it is consistent for different planetary masses.  The driving force of the increased instability at higher multiplicity is probably mainly a consequence of the intrinsically lower stability of high-multiplicity systems \citep{chambers1996,funk2010}, which is found both in previous works without mass loss and in our own simulations when we look at stability along the MS before mass loss. We speculate that this is a result of a larger number of multiple-planet resonances. However, non-adiabatic expansion may play a role for the systems of more massive planets where orbital radii can be large.

We also  find that a different dynamical behaviour is observed on the WD phase depending on the planet mass.
Here, we refer to planet masses of 1 and 10 $\mathrm{M_\oplus}$ as low-mass planets and planet masses of 100 and 1000 $\mathrm{M_\oplus}$ as high-mass planets. Simulations having orbit crossing without planet losses in the 10 Gyr tested are mainly restricted to systems involving low-mass planets. In addition, low-mass planets experience close encounters which are not strong enough to trigger a planet loss. The same dynamical outcome is found by \citet{veras2016b,mustill2018} in their  simulations of terrestrial planet mass. Indeed, the planet losses in our 1 $\mathrm{M_\oplus}$ simulations are restricted only to planet--planet collisions in all the multiple-planet configurations (see as well e.g. \citet{veras2013b,veras2015}). In addition, planet--star collisions happen more frequently than planet--planet collisions but much less frequently than ejections, and the higher the mass of the planets, the smaller the number of collisions with the star obtained. 

On the other hand, simulations involving high-mass planets preferably lose planets and  typically more than one planet is lost in a single simulation in the five- and six-planet systems (see Figs. \ref{pie14}, \ref{pie17}, \ref{pie30}). We point out that high-mass systems involve close encounters among the planets in which the orbit crossing strongly induces a planet loss afterwards, mainly by ejection.  In our simulations, the largest fractions of ejections occur in the systems with high-mass planets being the six-planet systems  with 1000 $\mathrm{M_\oplus}$ planets where $\sim$ 60 $\%$ of the planets are ejected for $\Delta\leq$ 14, 17. This tendency of a higher rate of ejections for higher mass planets has been previously reported by simulations in the literature (see e.g. \citealt{veras2013,veras2013b,mustill2014,mustill2018,veras2015,veras2016b}; \citetalias{maldonado2020}; \citetalias{maldonado2020b}; \citetalias{maldonado2021}) and it is well documented analytically by the Safronov number $\Theta$ c.f. \citealt{safronov1969,mustill2014}) that states that it is more likely that a planet scatter out other bodies as the rate of the escape velocity with the orbital velocity increases. The Safronov number using the mass of the WD and the innermost planet's orbit after the adiabatic expansion due to the mass-loss,   is 3.8, 12.1, 30.9, 290.0 for the four planet masses (1, 10, 100 and 1000 $\mathrm{M_\oplus}$) respectively. Thus,  our numerical results confirm the expected scatter out of more massive bodies as the planet/star mass ratio increases.

Furthermore, we find that simulations with high-mass planets tend to have the first orbit crossing at a WD cooling time $<$ 100 Myr, with the planet losses starting a few Myr later. The behaviour is slightly different for low-mass planets that have the first orbit crossing at cooling times $>$ 100 Myr and the first planet losses happen a few Gyr afterwards\footnote{We have calculated the geometric mean of four samples having the instability times converted to the WD's cooling time, two of them (1 and 10 $\mathrm{M_\oplus}$ simulations together and the other with 100 and 1000 $\mathrm{M_\oplus}$ together) with the information of the first orbit crossing in all the multiple-planet simulations dynamically active on the WD phase. The other two samples have the time when a planet is lost or when the first planet loss happens in case that multiple planets are lost in the same simulation. The geometric means for orbit crossing: high-mass -- 68 Myr, low-mass -- 351 Myr; planet losses: high-mass -- 331 Myr, low-mass -- 2869 Myr.}. Previous studies have described a similar tendency on the instability times between low-mass and high-mass planets (e.g. \citealt{veras2013,mustill2014}; \citetalias{maldonado2020b}; \citetalias{maldonado2021}) and \citet{veras2016b,mustill2018} also concluded that planets from 1 to 30 $\mathrm{M_\oplus}$ kept being dynamically active for Gyr time-scales.  We should note that although we have stated that at high planet masses, one or more planets are rapidly ejected, this does not necessarily lead to long term WD pollution since simulations that include asteroids show that the resulting accretion rate decays much more quickly (e.g. \citealt{mustill2018}), and they will produce a burst of accretion but little long-term delivery.

It can be clearly seen in Figs.\ \ref{pmass} and \ref{totwd} that the fraction of unstable simulations on the WD phase decreases as we consider systems with a higher planet separation $\Delta$, i.e. the fraction of dynamically active simulations is higher in systems with $\Delta\leq$ 14 than in systems with $\Delta\leq$ 17 and 30, respectively. This is an expected result given that it is the planet separation that provides the Hill and Lagrange stability limits at which planets are separated enough to have a dynamical evolution completely ordered and stable. This behaviour is also seen by comparing the fraction of stable systems obtained in our simulations in Figs. \ref{pie14}, \ref{pie17}, \ref{pie30}, which increases as $\Delta$ also increases.  In general, we find that the required planet separation $\Delta$ to have fully stable systems during the entire simulated time increases when the system has more planets, in agreement with \citep{funk2010}. The stability limits in our multiple-planet simulations are in agreement with the limits found in previous studies with two planets \citep{veras2013,veras2013b}, three planets \citep{mustill2014,mustill2018} and four planets \citep{veras2016b} by comparing the equivalent planetary masses and architectures.  Although we can certainly know when a pair of planets are prone to have close encounters by the Hill stability formalism \citep{hill1878,donnison2011}, there is not any analytical formulation to predict when a planetary system might be Lagrange unstable and it can only be known by performing numerical simulations.

\subsection{Links to different approaches}

 We highlight that orbital distances scaled close to planet mass to the 1/4th power should better suit the stability criterion in multiple-planet systems than planet distances scaled to planet masses to the 1/3rd power, as the Hill radius unit (see e.g. \citealt{petit2020,tamayo2021,rath2021}). Therefore, in order to see if there is a significant difference in the results adopting one or the other scaling, we compare the fractions of unstable simulations on the WD phase that we have found within the threshold distances in mutual Hill radius with respect to their corresponding fractions found within their equivalent thresholds that scale to 1/4th power of the planet mass \footnote{We have transformed our planet distances $\Delta$ in mutual Hill radius units to dynamical distances defined by equation 6 in \citet{tamayo2021}.}. We find small, but not significant, differences in the number of unstable simulations when comparing both scalings in their respective threshold distances; the general tendency of increasing fraction of unstable systems with the number of planets is kept. Perhaps with more simulations a significant difference might be found when comparing both scalings; however, it is hidden within the error bars in our sample.

We note that we have chosen the eccentricity from the same Rayleigh distribution  with $\sigma$ = 0.03 for all the planet masses. As a consequence, as the planet mass increases, so does the planet separation, thus, the eccentricity will become a fraction of the orbit crossing eccentricity. A caveat of this choice is that systems involving lower mass planets are, a priori, more likely to be dynamically unstable than systems having higher mass planets with the same planet separation. We defer to a future work  to  build a more dynamically comparable sample in terms of eccentricity, that is to re-scale the eccentricity with respect to the 10 $\mathrm{M_\oplus}$ simulations to be $\sigma_e$ $\sim$ 0.03.

 It is important to identify the link between the results obtained in this study and what is expected to be observed in real multiple-planet systems. However, we can not make a direct comparison because the real number of systems with multiple planets is hard to know \citep{johansen2012,tremaine2012,zhu2018,zink2019}, and it is likely that the observed planetary systems with one or two planets have additional planets but they remain hidden due to the planetary architectures \citep[for example, high mutual inclinations:][]{lissauer2011, fang2012,fabrycky2014,ballard2016}. It is important to note however that it has been estimated that the number of planets per G and K star is 5.86 $\pm$ 0.18 as a correction to the number of planets in the {\it Kepler} parameter space \citep{zink2019}. An additional reason why we can not compare our results with observed multiple-planet systems is due to the fact that the majority of these systems have all their planets located at semimajor axes $<$ 0.5 au and have sizes between Super-Earth to Sub-Neptune \citep{pu2015,weiss2018}. Only very few observed systems have planets with orbital distances comparable to the ones studied in this work, such as the planetary system HR~8799 with four giant planets \citep{marois2008,marois2010}. Moreover,  the observed multiple-planet systems have a mix of planet masses, eccentricities, planet separations \citep{schneider2011,akeson2013} and the planetary multiplicity is entangled with this mix in the physical and orbital parameter space of real systems, as it is shown in \citetalias{maldonado2020,maldonado2020b,maldonado2021}.  

 Systems may also provide pollution to the WD even if they do not undergo instability, for example through secular chaos \citep{connor2021}. The increase in orbital eccentricities we find going from the MS to the WD will enhance such secular chaos. The mechanism will also be more significant in high-multiplicity systems where secular resonances are more numerous. Thus, our approach of looking at direct instability among the planets gives a conservative estimate of the fraction of multiple-planet systems that may be responsible for WD pollution.

Finally, we point out that a hint about the trend of finding more dynamical instabilities with more planets in a planetary system was found in a more limited analysis by \citet{veras2015,veras2016b} when comparing their four-planet unstable simulations with the few tests involving six-, eight- and ten-planet systems. Furthermore, the fractions of dynamically active simulations in this work, regarding the two-, three- and four-planet systems with $\Delta\leq$ 14 on the WD phase, are comparable to the fractions found by \citet{veras2013,veras2013b,mustill2014,mustill2018,veras2015}, respectively.

\section{Conclusions}

 In this work we have numerically integrated the orbits of 5310 multiple-planet systems 
from the MS to the WD phase of a 3 $\mathrm{M_\odot}$ host star. The systems were built using different multiplicity configurations from two to six planets in which we set equal-mass planets with  four planet masses (1, 10, 100 and 1000 $\mathrm{M_\oplus}$) per planetary multiplicity, spacing the planets the same distance $\Delta$ from 2 to 30 mutual Hill radii per planet mass and adopting small eccentricities and orbital inclinations in all the templates. 

The main result of this study, consistent for different planet masses and invariant in the planet separation $\Delta$ points out that the fraction of unstable simulations on the WD phase (considering planets losses, orbit crossing and orbital scattering) in multiple-planet systems that share the same characteristics in the physical and orbital parameter space, increases as does the number of planets in the planetary system. We thus confirm the main conclusion of \citetalias{maldonado2021} in which scaled versions of observed multiple-planet systems were simulated and where the planetary multiplicity was part of the mixed parameter space. 

We find that the planet mass affects the dynamical evolution of multiple-planet systems and different outcomes are obtained, such as having a higher rate of planet--planet collisions and orbit crossing without planet losses in systems with low-mass (1 and 10 $\mathrm{M_\oplus}$) planets with respect to their high-mass (100 and 1000 $\mathrm{M_\oplus}$) counterpart. Moreover, ejections are more likely to happen in 
systems with high-mass planets, in agreement with the analytical Safronov number; conversely, planet--star collisions are less frequent as the mass of the planet increases. Furthermore, planet losses happen at a higher rate in systems with high-mass planets, and more planets are lost per planetary multiplicity in comparison to the low-mass simulations. On the other hand, the instability time in terms of the WD's cooling time span Gyr time-scales between the first orbit crossing and the planet losses in simulations with low-mass planets, while the simulations involving high-mass planets tend to become unstable in a few Myr after the WD phase starts.   

In our simulations, we obtain that the fraction of dynamically active simulations on the WD phase is higher as the planet separation $\Delta$ threshold  decreases. The wider the range of initial planet separations, the more stable systems are found, confirming the dependence of the stability limits with $\Delta$.  However, we also find that for a high multiplicity of planets (four, five and six planets), non-adiabatic effects are prone to appear at large $\Delta$ values, specially in the simulations with high-mass planets, causing the eccentricity excitation of the outer planets during the mass loss of the star, inducing in some cases more instabilities during the WD phase or even producing the imminent ejection of the outermost planets during the mass loss of the stellar host.

Multiple-planet system that are stable during the MS and survive the RGB and AGB phases of their stellar host may become dynamically active on the WD phase. We prove that the major role in increasing the rate of dynamical instabilities is played by the number of planets in the system. These instabilities may destabilize putative asteroid belts within the planetary systems, sending their minor components toward the central star which may cause several observed phenomena such as metal pollution on WD's atmospheres.

\section*{Acknowledgements}
 
  We thank to the referee for very insightful comments in the revision of the manuscript.     R.F.M. thanks the funding from the 'Collaboration Scholarship' granted by INAOE. R.F.M and E.V. acknowledge support from the `On the rocks II project' funded by the Spanish Ministerio de Ciencia, Innovaci\'on y Universidades under grant PGC2018-101950-B-I00 and the Unidad de Excelencia “María de Maeztu”- Centro de Astrobiología (CSIC/INTA).  M.C. thanks CONACyT for financial support through grant CB-2015-256961. A.J.M. acknowledges support from the starting grant 2017-04945 `A unified picture of white dwarf planetary systems' from the Swedish Research Council. The authors thankfully acknowledge the computer resources, technical expertise and support provided by the Laboratorio Nacional de Superc\'omputo del Sureste de M\'exico, CONACyT member of the network of national laboratories.

\section*{Data Availability}

The data underlying this article will be shared on reasonable request to the corresponding author.



\bibliographystyle{mnras}
\bibliography{bib} 

\begin{thebibliography}{}
\makeatletter
\relax
\def\mn@urlcharsother{\let\do\@makeother \do\$\do\&\do\#\do\^\do\_\do\%\do\~}
\def\mn@doi{\begingroup\mn@urlcharsother \@ifnextchar [ {\mn@doi@}
  {\mn@doi@[]}}
\def\mn@doi@[#1]#2{\def\@tempa{#1}\ifx\@tempa\@empty \href
  {http://dx.doi.org/#2} {doi:#2}\else \href {http://dx.doi.org/#2} {#1}\fi
  \endgroup}
\def\mn@eprint#1#2{\mn@eprint@#1:#2::\@nil}
\def\mn@eprint@arXiv#1{\href {http://arxiv.org/abs/#1} {{\tt arXiv:#1}}}
\def\mn@eprint@dblp#1{\href {http://dblp.uni-trier.de/rec/bibtex/#1.xml}
  {dblp:#1}}
\def\mn@eprint@#1:#2:#3:#4\@nil{\def\@tempa {#1}\def\@tempb {#2}\def\@tempc
  {#3}\ifx \@tempc \@empty \let \@tempc \@tempb \let \@tempb \@tempa \fi \ifx
  \@tempb \@empty \def\@tempb {arXiv}\fi \@ifundefined
  {mn@eprint@\@tempb}{\@tempb:\@tempc}{\expandafter \expandafter \csname
  mn@eprint@\@tempb\endcsname \expandafter{\@tempc}}}

\bibitem[\protect\citeauthoryear{{Akeson} et~al.,}{{Akeson}
  et~al.}{2013}]{akeson2013}
{Akeson} R.~L.,  et~al., 2013, \mn@doi [\pasp] {10.1086/672273}, \href
  {https://ui.adsabs.harvard.edu/abs/2013PASP..125..989A} {125, 989}

\bibitem[\protect\citeauthoryear{{Andrews}}{{Andrews}}{2020}]{sean2020}
{Andrews} S.~M.,  2020, \mn@doi [\araa] {10.1146/annurev-astro-031220-010302},
  \href {https://ui.adsabs.harvard.edu/abs/2020ARA&A..58..483A} {58, 483}

\bibitem[\protect\citeauthoryear{{Ballard} \& {Johnson}}{{Ballard} \&
  {Johnson}}{2016}]{ballard2016}
{Ballard} S.,  {Johnson} J.~A.,  2016, \mn@doi [\apj]
  {10.3847/0004-637X/816/2/66}, \href
  {https://ui.adsabs.harvard.edu/abs/2016ApJ...816...66B} {816, 66}

\bibitem[\protect\citeauthoryear{{Becklin}, {Farihi}, {Jura}, {Song},
  {Weinberger}  \& {Zuckerman}}{{Becklin} et~al.}{2005}]{becklin2005}
{Becklin} E.~E.,  {Farihi} J.,  {Jura} M.,  {Song} I.,  {Weinberger} A.~J.,
  {Zuckerman} B.,  2005, \mn@doi [\apjl] {10.1086/497826}, \href
  {https://ui.adsabs.harvard.edu/abs/2005ApJ...632L.119B} {632, L119}

\bibitem[\protect\citeauthoryear{{Blackman} et~al.,}{{Blackman}
  et~al.}{2021}]{blackman2021}
{Blackman} J.~W.,  et~al., 2021, \mn@doi [\nat] {10.1038/s41586-021-03869-6},
  \href {https://ui.adsabs.harvard.edu/abs/2021Natur.598..272B} {598, 272}

\bibitem[\protect\citeauthoryear{{Bonsor}, {Mustill}  \& {Wyatt}}{{Bonsor}
  et~al.}{2011}]{bonsor2011}
{Bonsor} A.,  {Mustill} A.~J.,   {Wyatt} M.~C.,  2011, \mn@doi [\mnras]
  {10.1111/j.1365-2966.2011.18524.x}, \href
  {http://adsabs.harvard.edu/abs/2011MNRAS.414..930B} {414, 930}

\bibitem[\protect\citeauthoryear{{Chambers}}{{Chambers}}{1999}]{chambers1999}
{Chambers} J.~E.,  1999, \mn@doi [\mnras] {10.1046/j.1365-8711.1999.02379.x},
  \href {http://adsabs.harvard.edu/abs/1999MNRAS.304..793C} {304, 793}

\bibitem[\protect\citeauthoryear{{Chambers}, {Wetherill}  \& {Boss}}{{Chambers}
  et~al.}{1996}]{chambers1996}
{Chambers} J.~E.,  {Wetherill} G.~W.,   {Boss} A.~P.,  1996, \mn@doi [\icarus]
  {10.1006/icar.1996.0019}, \href
  {https://ui.adsabs.harvard.edu/abs/1996Icar..119..261C} {119, 261}

\bibitem[\protect\citeauthoryear{{Chen} \& {Kipping}}{{Chen} \&
  {Kipping}}{2017}]{chen2017}
{Chen} J.,  {Kipping} D.,  2017, \mn@doi [\apj] {10.3847/1538-4357/834/1/17},
  \href {http://adsabs.harvard.edu/abs/2017ApJ...834...17C} {834, 17}

\bibitem[\protect\citeauthoryear{{Debes} \& {Sigurdsson}}{{Debes} \&
  {Sigurdsson}}{2002}]{debes2002}
{Debes} J.~H.,  {Sigurdsson} S.,  2002, \mn@doi [\apj] {10.1086/340291}, \href
  {http://adsabs.harvard.edu/abs/2002ApJ...572..556D} {572, 556}

\bibitem[\protect\citeauthoryear{{Debes}, {Walsh}  \& {Stark}}{{Debes}
  et~al.}{2012}]{debes2012}
{Debes} J.~H.,  {Walsh} K.~J.,   {Stark} C.,  2012, \mn@doi [\apj]
  {10.1088/0004-637X/747/2/148}, \href
  {http://adsabs.harvard.edu/abs/2012ApJ...747..148D} {747, 148}

\bibitem[\protect\citeauthoryear{{Dennihy} et~al.,}{{Dennihy}
  et~al.}{2020}]{dennihy2020}
{Dennihy} E.,  et~al., 2020, \mn@doi [\apj] {10.3847/1538-4357/abc339}, \href
  {https://ui.adsabs.harvard.edu/abs/2020ApJ...905....5D} {905, 5}

\bibitem[\protect\citeauthoryear{{Donnison}}{{Donnison}}{2011}]{donnison2011}
{Donnison} J.~R.,  2011, \mn@doi [\mnras] {10.1111/j.1365-2966.2011.18720.x},
  \href {http://adsabs.harvard.edu/abs/2011MNRAS.415..470D} {415, 470}

\bibitem[\protect\citeauthoryear{{Fabrycky} et~al.,}{{Fabrycky}
  et~al.}{2014}]{fabrycky2014}
{Fabrycky} D.~C.,  et~al., 2014, \mn@doi [\apj] {10.1088/0004-637X/790/2/146},
  \href {https://ui.adsabs.harvard.edu/abs/2014ApJ...790..146F} {790, 146}

\bibitem[\protect\citeauthoryear{{Fang} \& {Margot}}{{Fang} \&
  {Margot}}{2012}]{fang2012}
{Fang} J.,  {Margot} J.-L.,  2012, \mn@doi [\apj] {10.1088/0004-637X/761/2/92},
  \href {https://ui.adsabs.harvard.edu/abs/2012ApJ...761...92F} {761, 92}

\bibitem[\protect\citeauthoryear{{Farihi} et~al.,}{{Farihi}
  et~al.}{2022}]{farihi2022}
{Farihi} J.,  et~al., 2022, \mn@doi [\mnras] {10.1093/mnras/stab3475}, \href
  {https://ui.adsabs.harvard.edu/abs/2022MNRAS.511.1647F} {511, 1647}

\bibitem[\protect\citeauthoryear{{Frewen} \& {Hansen}}{{Frewen} \&
  {Hansen}}{2014}]{frewen2014}
{Frewen} S.~F.~N.,  {Hansen} B.~M.~S.,  2014, \mn@doi [\mnras]
  {10.1093/mnras/stu097}, \href
  {http://adsabs.harvard.edu/abs/2014MNRAS.439.2442F} {439, 2442}

\bibitem[\protect\citeauthoryear{{Funk}, {Wuchterl}, {Schwarz},
  {Pilat-Lohinger}  \& {Eggl}}{{Funk} et~al.}{2010}]{funk2010}
{Funk} B.,  {Wuchterl} G.,  {Schwarz} R.,  {Pilat-Lohinger} E.,   {Eggl} S.,
  2010, \mn@doi [\aap] {10.1051/0004-6361/200912698}, \href
  {https://ui.adsabs.harvard.edu/abs/2010A&A...516A..82F} {516, A82}

\bibitem[\protect\citeauthoryear{{G{\"a}nsicke}, {Marsh}, {Southworth}  \&
  {Rebassa-Mansergas}}{{G{\"a}nsicke} et~al.}{2006}]{gansicke2006}
{G{\"a}nsicke} B.~T.,  {Marsh} T.~R.,  {Southworth} J.,   {Rebassa-Mansergas}
  A.,  2006, \mn@doi [Science] {10.1126/science.1135033}, \href
  {http://adsabs.harvard.edu/abs/2006Sci...314.1908G} {314, 1908}

\bibitem[\protect\citeauthoryear{{G{\"a}nsicke}, {Schreiber}, {Toloza},
  {Fusillo}, {Koester}  \& {Manser}}{{G{\"a}nsicke}
  et~al.}{2019}]{gansicke2019}
{G{\"a}nsicke} B.~T.,  {Schreiber} M.~R.,  {Toloza} O.,  {Fusillo} N. P.~G.,
  {Koester} D.,   {Manser} C.~J.,  2019, \mn@doi [\nat]
  {10.1038/s41586-019-1789-8}, \href
  {https://ui.adsabs.harvard.edu/abs/2019Natur.576...61G} {576, 61}

\bibitem[\protect\citeauthoryear{{Garufi} et~al.,}{{Garufi}
  et~al.}{2020}]{garufi2020}
{Garufi} A.,  et~al., 2020, \mn@doi [\aap] {10.1051/0004-6361/201936946}, \href
  {https://ui.adsabs.harvard.edu/abs/2020A&A...633A..82G} {633, A82}

\bibitem[\protect\citeauthoryear{{Guidry} et~al.,}{{Guidry}
  et~al.}{2021}]{guidry2021}
{Guidry} J.~A.,  et~al., 2021, \mn@doi [\apj] {10.3847/1538-4357/abee68}, \href
  {https://ui.adsabs.harvard.edu/abs/2021ApJ...912..125G} {912, 125}

\bibitem[\protect\citeauthoryear{{Hill}}{{Hill}}{1878}]{hill1878}
{Hill} G.~W.,  1878, Am. J. Math., 1, 129

\bibitem[\protect\citeauthoryear{{Hurley}, {Pols}  \& {Tout}}{{Hurley}
  et~al.}{2000}]{hurley2000}
{Hurley} J.~R.,  {Pols} O.~R.,   {Tout} C.~A.,  2000, \mn@doi [\mnras]
  {10.1046/j.1365-8711.2000.03426.x}, \href
  {http://adsabs.harvard.edu/abs/2000MNRAS.315..543H} {315, 543}

\bibitem[\protect\citeauthoryear{{Jaynes} \& {Bretthorst}}{{Jaynes} \&
  {Bretthorst}}{2003}]{jaynes2003}
{Jaynes} E.~T.,  {Bretthorst} G.~L.,  2003, {Probability Theory}

\bibitem[\protect\citeauthoryear{{Johansen}, {Davies}, {Church}  \&
  {Holmelin}}{{Johansen} et~al.}{2012}]{johansen2012}
{Johansen} A.,  {Davies} M.~B.,  {Church} R.~P.,   {Holmelin} V.,  2012,
  \mn@doi [\apj] {10.1088/0004-637X/758/1/39}, \href
  {https://ui.adsabs.harvard.edu/abs/2012ApJ...758...39J} {758, 39}

\bibitem[\protect\citeauthoryear{{Kalirai}, {Hansen}, {Kelson}, {Reitzel},
  {Rich}  \& {Richer}}{{Kalirai} et~al.}{2008}]{kalirai2008}
{Kalirai} J.~S.,  {Hansen} B.~M.~S.,  {Kelson} D.~D.,  {Reitzel} D.~B.,  {Rich}
  R.~M.,   {Richer} H.~B.,  2008, \mn@doi [\apj] {10.1086/527028}, \href
  {http://adsabs.harvard.edu/abs/2008ApJ...676..594K} {676, 594}

\bibitem[\protect\citeauthoryear{{Kilic} \& {Redfield}}{{Kilic} \&
  {Redfield}}{2007}]{kilic2007}
{Kilic} M.,  {Redfield} S.,  2007, \mn@doi [\apj] {10.1086/513008}, \href
  {http://adsabs.harvard.edu/abs/2007ApJ...660..641K} {660, 641}

\bibitem[\protect\citeauthoryear{{Koester}, {G{\"a}nsicke}  \&
  {Farihi}}{{Koester} et~al.}{2014}]{koester2014}
{Koester} D.,  {G{\"a}nsicke} B.~T.,   {Farihi} J.,  2014, \mn@doi [\aap]
  {10.1051/0004-6361/201423691}, \href
  {http://adsabs.harvard.edu/abs/2014A%26A...566A..34K} {566, A34}

\bibitem[\protect\citeauthoryear{{Kunitomo}, {Ikoma}, {Sato}, {Katsuta}  \&
  {Ida}}{{Kunitomo} et~al.}{2011}]{kunitomo2011}
{Kunitomo} M.,  {Ikoma} M.,  {Sato} B.,  {Katsuta} Y.,   {Ida} S.,  2011,
  \mn@doi [\apj] {10.1088/0004-637X/737/2/66}, \href
  {https://ui.adsabs.harvard.edu/abs/2011ApJ...737...66K} {737, 66}

\bibitem[\protect\citeauthoryear{{Li}, {Mustill}  \& {Davies}}{{Li}
  et~al.}{2021}]{li2021}
{Li} D.,  {Mustill} A.~J.,   {Davies} M.~B.,  2021, \mn@doi [\mnras]
  {10.1093/mnras/stab2949}, \href
  {https://ui.adsabs.harvard.edu/abs/2021MNRAS.508.5671L} {508, 5671}

\bibitem[\protect\citeauthoryear{{Lissauer} et~al.,}{{Lissauer}
  et~al.}{2011}]{lissauer2011}
{Lissauer} J.~J.,  et~al., 2011, \mn@doi [\apjs] {10.1088/0067-0049/197/1/8},
  \href {https://ui.adsabs.harvard.edu/abs/2011ApJS..197....8L} {197, 8}

\bibitem[\protect\citeauthoryear{{Maldonado}, {Villaver}, {Mustill}, {Chavez}
  \& {Bertone}}{{Maldonado} et~al.}{2020a}]{maldonado2020}
{Maldonado} R.~F.,  {Villaver} E.,  {Mustill} A.~J.,  {Chavez} M.,   {Bertone}
  E.,  2020a, \mn@doi [\mnras] {10.1093/mnras/staa2237}, \href
  {https://ui.adsabs.harvard.edu/abs/2020MNRAS.497.4091M} {497, 4091}

\bibitem[\protect\citeauthoryear{{Maldonado}, {Villaver}, {Mustill}, {Chavez}
  \& {Bertone}}{{Maldonado} et~al.}{2020b}]{maldonado2020b}
{Maldonado} R.~F.,  {Villaver} E.,  {Mustill} A.~J.,  {Chavez} M.,   {Bertone}
  E.,  2020b, \mn@doi [\mnras] {10.1093/mnras/staa2946}, \href
  {https://ui.adsabs.harvard.edu/abs/2020MNRAS.499.1854M} {499, 1854}

\bibitem[\protect\citeauthoryear{{Maldonado}, {Villaver}, {Mustill},
  {Ch{\'a}vez}  \& {Bertone}}{{Maldonado} et~al.}{2021}]{maldonado2021}
{Maldonado} R.~F.,  {Villaver} E.,  {Mustill} A.~J.,  {Ch{\'a}vez} M.,
  {Bertone} E.,  2021, \mn@doi [\mnras] {10.1093/mnrasl/slaa193}, \href
  {https://ui.adsabs.harvard.edu/abs/2021MNRAS.501L..43M} {501, L43}

\bibitem[\protect\citeauthoryear{{Manser}, {G{\"a}nsicke}, {Koester}, {Marsh}
  \& {Southworth}}{{Manser} et~al.}{2016}]{manser2016}
{Manser} C.~J.,  {G{\"a}nsicke} B.~T.,  {Koester} D.,  {Marsh} T.~R.,
  {Southworth} J.,  2016, \mn@doi [\mnras] {10.1093/mnras/stw1760}, \href
  {http://adsabs.harvard.edu/abs/2016MNRAS.462.1461M} {462, 1461}

\bibitem[\protect\citeauthoryear{{Marois}, {Macintosh}, {Barman}, {Zuckerman},
  {Song}, {Patience}, {Lafreni{\`e}re}  \& {Doyon}}{{Marois}
  et~al.}{2008}]{marois2008}
{Marois} C.,  {Macintosh} B.,  {Barman} T.,  {Zuckerman} B.,  {Song} I.,
  {Patience} J.,  {Lafreni{\`e}re} D.,   {Doyon} R.,  2008, \mn@doi [Science]
  {10.1126/science.1166585}, \href
  {https://ui.adsabs.harvard.edu/abs/2008Sci...322.1348M} {322, 1348}

\bibitem[\protect\citeauthoryear{{Marois}, {Zuckerman}, {Konopacky},
  {Macintosh}  \& {Barman}}{{Marois} et~al.}{2010}]{marois2010}
{Marois} C.,  {Zuckerman} B.,  {Konopacky} Q.~M.,  {Macintosh} B.,   {Barman}
  T.,  2010, \mn@doi [\nat] {10.1038/nature09684}, \href
  {https://ui.adsabs.harvard.edu/abs/2010Natur.468.1080M} {468, 1080}

\bibitem[\protect\citeauthoryear{{Melis}, {Klein}, {Doyle}, {Weinberger},
  {Zuckerman}  \& {Dufour}}{{Melis} et~al.}{2020}]{melis2020}
{Melis} C.,  {Klein} B.,  {Doyle} A.~E.,  {Weinberger} A.,  {Zuckerman} B.,
  {Dufour} P.,  2020, \mn@doi [\apj] {10.3847/1538-4357/abbdfa}, \href
  {https://ui.adsabs.harvard.edu/abs/2020ApJ...905...56M} {905, 56}

\bibitem[\protect\citeauthoryear{{Moorhead} et~al.,}{{Moorhead}
  et~al.}{2011}]{moorhead2011}
{Moorhead} A.~V.,  et~al., 2011, \mn@doi [\apjs] {10.1088/0067-0049/197/1/1},
  \href {http://adsabs.harvard.edu/abs/2011ApJS..197....1M} {197, 1}

\bibitem[\protect\citeauthoryear{{Mustill} \& {Villaver}}{{Mustill} \&
  {Villaver}}{2012}]{mustill2012}
{Mustill} A.~J.,  {Villaver} E.,  2012, \mn@doi [\apj]
  {10.1088/0004-637X/761/2/121}, \href
  {http://adsabs.harvard.edu/abs/2012ApJ...761..121M} {761, 121}

\bibitem[\protect\citeauthoryear{{Mustill} \& {Wyatt}}{{Mustill} \&
  {Wyatt}}{2012}]{mustill2012b}
{Mustill} A.~J.,  {Wyatt} M.~C.,  2012, \mn@doi [\mnras]
  {10.1111/j.1365-2966.2011.19948.x}, \href
  {https://ui.adsabs.harvard.edu/abs/2012MNRAS.419.3074M} {419, 3074}

\bibitem[\protect\citeauthoryear{{Mustill}, {Veras}  \& {Villaver}}{{Mustill}
  et~al.}{2014}]{mustill2014}
{Mustill} A.~J.,  {Veras} D.,   {Villaver} E.,  2014, \mn@doi [\mnras]
  {10.1093/mnras/stt1973}, \href
  {http://adsabs.harvard.edu/abs/2014MNRAS.437.1404M} {437, 1404}

\bibitem[\protect\citeauthoryear{{Mustill}, {Villaver}, {Veras}, {G{\"a}nsicke}
   \& {Bonsor}}{{Mustill} et~al.}{2018}]{mustill2018}
{Mustill} A.~J.,  {Villaver} E.,  {Veras} D.,  {G{\"a}nsicke} B.~T.,   {Bonsor}
  A.,  2018, \mn@doi [\mnras] {10.1093/mnras/sty446}, \href
  {http://adsabs.harvard.edu/abs/2018MNRAS.476.3939M} {476, 3939}

\bibitem[\protect\citeauthoryear{{Nordhaus} \& {Spiegel}}{{Nordhaus} \&
  {Spiegel}}{2013}]{nordhaus2013}
{Nordhaus} J.,  {Spiegel} D.~S.,  2013, \mn@doi [\mnras]
  {10.1093/mnras/stt569}, \href
  {https://ui.adsabs.harvard.edu/abs/2013MNRAS.432..500N} {432, 500}

\bibitem[\protect\citeauthoryear{{O'Connor}, {Teyssandier}  \&
  {Lai}}{{O'Connor} et~al.}{2021}]{connor2021}
{O'Connor} C.~E.,  {Teyssandier} J.,   {Lai} D.,  2021, arXiv e-prints, \href
  {https://ui.adsabs.harvard.edu/abs/2021arXiv211108716O} {p. arXiv:2111.08716}

\bibitem[\protect\citeauthoryear{{Petit}, {Pichierri}, {Davies}  \&
  {Johansen}}{{Petit} et~al.}{2020}]{petit2020}
{Petit} A.~C.,  {Pichierri} G.,  {Davies} M.~B.,   {Johansen} A.,  2020,
  \mn@doi [\aap] {10.1051/0004-6361/202038764}, \href
  {https://ui.adsabs.harvard.edu/abs/2020A&A...641A.176P} {641, A176}

\bibitem[\protect\citeauthoryear{{Pu} \& {Wu}}{{Pu} \& {Wu}}{2015}]{pu2015}
{Pu} B.,  {Wu} Y.,  2015, \mn@doi [\apj] {10.1088/0004-637X/807/1/44}, \href
  {http://adsabs.harvard.edu/abs/2015ApJ...807...44P} {807, 44}

\bibitem[\protect\citeauthoryear{{Rath}, {Hadden}  \& {Lithwick}}{{Rath}
  et~al.}{2021}]{rath2021}
{Rath} J.,  {Hadden} S.,   {Lithwick} Y.,  2021, arXiv e-prints, \href
  {https://ui.adsabs.harvard.edu/abs/2021arXiv211002956R} {p. arXiv:2110.02956}

\bibitem[\protect\citeauthoryear{{Rebassa-Mansergas}, {Solano}, {Xu},
  {Rodrigo}, {Jim{\'e}nez-Esteban}  \& {Torres}}{{Rebassa-Mansergas}
  et~al.}{2019}]{rebassa2019}
{Rebassa-Mansergas} A.,  {Solano} E.,  {Xu} S.,  {Rodrigo} C.,
  {Jim{\'e}nez-Esteban} F.~M.,   {Torres} S.,  2019, \mn@doi [\mnras]
  {10.1093/mnras/stz2423}, \href
  {https://ui.adsabs.harvard.edu/abs/2019MNRAS.489.3990R} {489, 3990}

\bibitem[\protect\citeauthoryear{{Rocchetto}, {Farihi}, {G{\"a}nsicke}  \&
  {Bergfors}}{{Rocchetto} et~al.}{2015}]{rocchetto2015}
{Rocchetto} M.,  {Farihi} J.,  {G{\"a}nsicke} B.~T.,   {Bergfors} C.,  2015,
  \mn@doi [\mnras] {10.1093/mnras/stv282}, \href
  {https://ui.adsabs.harvard.edu/abs/2015MNRAS.449..574R} {449, 574}

\bibitem[\protect\citeauthoryear{{Ronco}, {Schreiber}, {Giuppone}, {Veras},
  {Cuadra}  \& {Guilera}}{{Ronco} et~al.}{2020}]{ronco2020}
{Ronco} M.~P.,  {Schreiber} M.~R.,  {Giuppone} C.~A.,  {Veras} D.,  {Cuadra}
  J.,   {Guilera} O.~M.,  2020, \mn@doi [\apjl] {10.3847/2041-8213/aba35f},
  \href {https://ui.adsabs.harvard.edu/abs/2020ApJ...898L..23R} {898, L23}

\bibitem[\protect\citeauthoryear{{Safronov} \& {Zvjagina}}{{Safronov} \&
  {Zvjagina}}{1969}]{safronov1969}
{Safronov} V.~S.,  {Zvjagina} E.~V.,  1969, \mn@doi [\icarus]
  {10.1016/0019-1035(69)90013-X}, \href
  {https://ui.adsabs.harvard.edu/abs/1969Icar...10..109S} {10, 109}

\bibitem[\protect\citeauthoryear{{Schneider}, {Dedieu}, {Le Sidaner}, {Savalle}
   \& {Zolotukhin}}{{Schneider} et~al.}{2011}]{schneider2011}
{Schneider} J.,  {Dedieu} C.,  {Le Sidaner} P.,  {Savalle} R.,   {Zolotukhin}
  I.,  2011, \mn@doi [\aap] {10.1051/0004-6361/201116713}, \href
  {https://ui.adsabs.harvard.edu/abs/2011A&A...532A..79S} {532, A79}

\bibitem[\protect\citeauthoryear{{Smallwood}, {Martin}, {Livio}  \&
  {Lubow}}{{Smallwood} et~al.}{2018}]{smallwood2018}
{Smallwood} J.~L.,  {Martin} R.~G.,  {Livio} M.,   {Lubow} S.~H.,  2018,
  \mn@doi [\mnras] {10.1093/mnras/sty1819}, \href
  {https://ui.adsabs.harvard.edu/abs/2018MNRAS.480...57S} {480, 57}

\bibitem[\protect\citeauthoryear{{Tamayo}, {Murray}, {Tremaine}  \&
  {Winn}}{{Tamayo} et~al.}{2021}]{tamayo2021}
{Tamayo} D.,  {Murray} N.,  {Tremaine} S.,   {Winn} J.,  2021, \mn@doi [\aj]
  {10.3847/1538-3881/ac1c6a}, \href
  {https://ui.adsabs.harvard.edu/abs/2021AJ....162..220T} {162, 220}

\bibitem[\protect\citeauthoryear{{Tazzari} et~al.,}{{Tazzari}
  et~al.}{2017}]{tazzari2017}
{Tazzari} M.,  et~al., 2017, \mn@doi [\aap] {10.1051/0004-6361/201730890},
  \href {https://ui.adsabs.harvard.edu/abs/2017A&A...606A..88T} {606, A88}

\bibitem[\protect\citeauthoryear{{Tremaine} \& {Dong}}{{Tremaine} \&
  {Dong}}{2012}]{tremaine2012}
{Tremaine} S.,  {Dong} S.,  2012, \mn@doi [\aj] {10.1088/0004-6256/143/4/94},
  \href {https://ui.adsabs.harvard.edu/abs/2012AJ....143...94T} {143, 94}

\bibitem[\protect\citeauthoryear{{Van Eylen} \& {Albrecht}}{{Van Eylen} \&
  {Albrecht}}{2015}]{vaneylen2015}
{Van Eylen} V.,  {Albrecht} S.,  2015, \mn@doi [\apj]
  {10.1088/0004-637X/808/2/126}, \href
  {https://ui.adsabs.harvard.edu/abs/2015ApJ...808..126V} {808, 126}

\bibitem[\protect\citeauthoryear{{Vanderbosch} et~al.,}{{Vanderbosch}
  et~al.}{2020}]{vanderbosch2020}
{Vanderbosch} Z.,  et~al., 2020, \mn@doi [\apj] {10.3847/1538-4357/ab9649},
  \href {https://ui.adsabs.harvard.edu/abs/2020ApJ...897..171V} {897, 171}

\bibitem[\protect\citeauthoryear{{Vanderbosch} et~al.,}{{Vanderbosch}
  et~al.}{2021}]{vanderbosch2021b}
{Vanderbosch} Z.~P.,  et~al., 2021, \mn@doi [\apj] {10.3847/1538-4357/ac0822},
  \href {https://ui.adsabs.harvard.edu/abs/2021ApJ...917...41V} {917, 41}

\bibitem[\protect\citeauthoryear{{Vanderburg} et~al.,}{{Vanderburg}
  et~al.}{2015}]{vanderburg2015}
{Vanderburg} A.,  et~al., 2015, \mn@doi [\nat] {10.1038/nature15527}, \href
  {http://adsabs.harvard.edu/abs/2015Natur.526..546V} {526, 546}

\bibitem[\protect\citeauthoryear{{Vanderburg} et~al.,}{{Vanderburg}
  et~al.}{2020}]{vanderburg2020}
{Vanderburg} A.,  et~al., 2020, \mn@doi [\nat] {10.1038/s41586-020-2713-y},
  \href {https://ui.adsabs.harvard.edu/abs/2020Natur.585..363V} {585, 363}

\bibitem[\protect\citeauthoryear{{Veras} \& {Evans}}{{Veras} \&
  {Evans}}{2013}]{veras2013c}
{Veras} D.,  {Evans} N.~W.,  2013, \mn@doi [\mnras] {10.1093/mnras/sts647},
  \href {https://ui.adsabs.harvard.edu/abs/2013MNRAS.430..403V} {430, 403}

\bibitem[\protect\citeauthoryear{{Veras} \& {G{\"a}nsicke}}{{Veras} \&
  {G{\"a}nsicke}}{2015}]{veras2015}
{Veras} D.,  {G{\"a}nsicke} B.~T.,  2015, \mn@doi [\mnras]
  {10.1093/mnras/stu2475}, \href
  {http://adsabs.harvard.edu/abs/2015MNRAS.447.1049V} {447, 1049}

\bibitem[\protect\citeauthoryear{{Veras} \& {Mustill}}{{Veras} \&
  {Mustill}}{2013}]{veras2013}
{Veras} D.,  {Mustill} A.~J.,  2013, \mn@doi [\mnras] {10.1093/mnrasl/slt067},
  \href {http://adsabs.harvard.edu/abs/2013MNRAS.434L..11V} {434, L11}

\bibitem[\protect\citeauthoryear{{Veras}, {Wyatt}, {Mustill}, {Bonsor}  \&
  {Eldridge}}{{Veras} et~al.}{2011}]{veras2011}
{Veras} D.,  {Wyatt} M.~C.,  {Mustill} A.~J.,  {Bonsor} A.,   {Eldridge} J.~J.,
   2011, \mn@doi [\mnras] {10.1111/j.1365-2966.2011.19393.x}, \href
  {https://ui.adsabs.harvard.edu/abs/2011MNRAS.417.2104V} {417, 2104}

\bibitem[\protect\citeauthoryear{{Veras}, {Mustill}, {Bonsor}  \&
  {Wyatt}}{{Veras} et~al.}{2013}]{veras2013b}
{Veras} D.,  {Mustill} A.~J.,  {Bonsor} A.,   {Wyatt} M.~C.,  2013, \mn@doi
  [\mnras] {10.1093/mnras/stt289}, \href
  {http://adsabs.harvard.edu/abs/2013MNRAS.431.1686V} {431, 1686}

\bibitem[\protect\citeauthoryear{{Veras}, {Evans}, {Wyatt}  \& {Tout}}{{Veras}
  et~al.}{2014}]{veras2014c}
{Veras} D.,  {Evans} N.~W.,  {Wyatt} M.~C.,   {Tout} C.~A.,  2014, \mn@doi
  [\mnras] {10.1093/mnras/stt1905}, \href
  {https://ui.adsabs.harvard.edu/abs/2014MNRAS.437.1127V} {437, 1127}

\bibitem[\protect\citeauthoryear{{Veras}, {Mustill}, {G{\"a}nsicke},
  {Redfield}, {Georgakarakos}, {Bowler}  \& {Lloyd}}{{Veras}
  et~al.}{2016}]{veras2016b}
{Veras} D.,  {Mustill} A.~J.,  {G{\"a}nsicke} B.~T.,  {Redfield} S.,
  {Georgakarakos} N.,  {Bowler} A.~B.,   {Lloyd} M.~J.~S.,  2016, \mn@doi
  [\mnras] {10.1093/mnras/stw476}, \href
  {http://adsabs.harvard.edu/abs/2016MNRAS.458.3942V} {458, 3942}

\bibitem[\protect\citeauthoryear{{Veras}, {Georgakarakos}, {G{\"a}nsicke}  \&
  {Dobbs-Dixon}}{{Veras} et~al.}{2018}]{veras2018}
{Veras} D.,  {Georgakarakos} N.,  {G{\"a}nsicke} B.~T.,   {Dobbs-Dixon} I.,
  2018, \mn@doi [\mnras] {10.1093/mnras/sty2409}, \href
  {http://adsabs.harvard.edu/abs/2018MNRAS.481.2180V} {481, 2180}

\bibitem[\protect\citeauthoryear{{Veras}, {Georgakarakos}, {Mustill},
  {Malamud}, {Cunningham}  \& {Dobbs-Dixon}}{{Veras} et~al.}{2021}]{veras2021}
{Veras} D.,  {Georgakarakos} N.,  {Mustill} A.~J.,  {Malamud} U.,  {Cunningham}
  T.,   {Dobbs-Dixon} I.,  2021, \mn@doi [\mnras] {10.1093/mnras/stab1667},
  \href {https://ui.adsabs.harvard.edu/abs/2021MNRAS.506.1148V} {506, 1148}

\bibitem[\protect\citeauthoryear{{Villaver} \& {Livio}}{{Villaver} \&
  {Livio}}{2009}]{villaver2009}
{Villaver} E.,  {Livio} M.,  2009, \mn@doi [\apjl]
  {10.1088/0004-637X/705/1/L81}, \href
  {http://adsabs.harvard.edu/abs/2009ApJ...705L..81V} {705, L81}

\bibitem[\protect\citeauthoryear{{Villaver}, {Livio}, {Mustill}  \&
  {Siess}}{{Villaver} et~al.}{2014}]{villaver2014}
{Villaver} E.,  {Livio} M.,  {Mustill} A.~J.,   {Siess} L.,  2014, \mn@doi
  [\apj] {10.1088/0004-637X/794/1/3}, \href
  {https://ui.adsabs.harvard.edu/abs/2014ApJ...794....3V} {794, 3}

\bibitem[\protect\citeauthoryear{{Voyatzis}, {Hadjidemetriou}, {Veras}  \&
  {Varvoglis}}{{Voyatzis} et~al.}{2013}]{voyatzis2013}
{Voyatzis} G.,  {Hadjidemetriou} J.~D.,  {Veras} D.,   {Varvoglis} H.,  2013,
  \mn@doi [\mnras] {10.1093/mnras/stt137}, \href
  {https://ui.adsabs.harvard.edu/abs/2013MNRAS.430.3383V} {430, 3383}

\bibitem[\protect\citeauthoryear{{Weiss} et~al.,}{{Weiss}
  et~al.}{2018}]{weiss2018}
{Weiss} L.~M.,  et~al., 2018, \mn@doi [\aj] {10.3847/1538-3881/aa9ff6}, \href
  {http://adsabs.harvard.edu/abs/2018AJ....155...48W} {155, 48}

\bibitem[\protect\citeauthoryear{{Xie} et~al.,}{{Xie} et~al.}{2016}]{xie2016}
{Xie} J.-W.,  et~al., 2016, \mn@doi [Proceedings of the National Academy of
  Science] {10.1073/pnas.1604692113}, \href
  {http://adsabs.harvard.edu/abs/2016PNAS..11311431X} {113, 11431}

\bibitem[\protect\citeauthoryear{{Zhu}, {Petrovich}, {Wu}, {Dong}  \&
  {Xie}}{{Zhu} et~al.}{2018}]{zhu2018}
{Zhu} W.,  {Petrovich} C.,  {Wu} Y.,  {Dong} S.,   {Xie} J.,  2018, \mn@doi
  [\apj] {10.3847/1538-4357/aac6d5}, \href
  {https://ui.adsabs.harvard.edu/abs/2018ApJ...860..101Z} {860, 101}

\bibitem[\protect\citeauthoryear{{Zink}, {Christiansen}  \& {Hansen}}{{Zink}
  et~al.}{2019}]{zink2019}
{Zink} J.~K.,  {Christiansen} J.~L.,   {Hansen} B. M.~S.,  2019, \mn@doi
  [\mnras] {10.1093/mnras/sty3463}, \href
  {https://ui.adsabs.harvard.edu/abs/2019MNRAS.483.4479Z} {483, 4479}

\bibitem[\protect\citeauthoryear{{Zink}, {Batygin}  \& {Adams}}{{Zink}
  et~al.}{2020}]{zinc2020}
{Zink} J.~K.,  {Batygin} K.,   {Adams} F.~C.,  2020, \mn@doi [\aj]
  {10.3847/1538-3881/abb8de}, \href
  {https://ui.adsabs.harvard.edu/abs/2020AJ....160..232Z} {160, 232}

\bibitem[\protect\citeauthoryear{{Zuckerman} \& {Becklin}}{{Zuckerman} \&
  {Becklin}}{1987}]{zuckerman1987}
{Zuckerman} B.,  {Becklin} E.~E.,  1987, \mn@doi [\nat] {10.1038/330138a0},
  \href {http://adsabs.harvard.edu/abs/1987Natur.330..138Z} {330, 138}

\bibitem[\protect\citeauthoryear{{Zuckerman}, {Koester}, {Reid}  \&
  {H{\"u}nsch}}{{Zuckerman} et~al.}{2003}]{zuckerman2003}
{Zuckerman} B.,  {Koester} D.,  {Reid} I.~N.,   {H{\"u}nsch} M.,  2003, \mn@doi
  [\apj] {10.1086/377492}, \href
  {https://ui.adsabs.harvard.edu/abs/2003ApJ...596..477Z} {596, 477}

\makeatother
\end{thebibliography}









\bsp	
\label{lastpage}
\end{document}